\documentclass[pre,twocolumn,nobalancelastpage,amsmath,amssymb,showpacs,
nofootinbib,superscriptaddress]{revtex4-1}

\usepackage{graphics}      
\usepackage{graphicx}      
\usepackage{longtable}     
\usepackage{url}           
\usepackage{bm}            
\usepackage{xcolor}
\usepackage{amsmath}
\usepackage{dcolumn}

\begin{document}
\title{Critical Scaling of Bagnold Rheology at the Jamming Transition of Frictionless Two Dimensional Disks}
\author{Daniel V{\aa}gberg}
\affiliation{Process \& Energy Laboratory, Delft University of Technology, Leeghwaterstraat 39, 2628 CB Delft, The Netherlands}
\author{Peter Olsson}
\affiliation{Department of Physics, Ume{\aa} University, 901 87 Ume{\aa}, Sweden}
\author{S. Teitel}
\affiliation{Department of Physics and Astronomy, University of Rochester, Rochester, NY 14627}
\date{\today}

\begin{abstract}
We carry out constant volume simulations of steady-state, shear driven, rheology in a simple model of bidisperse, soft-core, frictionless disks in two dimensions, using a dissipation law that gives rise to Bagnoldian rheology.  We discuss in detail the critical scaling ansatz for the shear-driven jamming transition, and carry out a detailed scaling analysis of our resulting data for pressure $p$ and shear stress $\sigma$.  Our analysis determines the critical exponent $\beta$ that describes the algebraic divergence of the Bagnold transport coefficients, $\lim_{\dot\gamma\to 0}p/\dot\gamma^2,\sigma/\dot\gamma^2\sim (\phi_J-\phi)^{-\beta}$, as the jamming transition $\phi_J$ is approached from below.  For the low strain rates considered in this work, we show that it is still necessary to consider the leading correction-to-scaling term in order to achieve a self-consistent analysis of our data, in which the critical parameters become independent of the size of the window of data used in the analysis.
We compare our resulting value  $\beta\approx 5.0\pm 0.4$ against previous numerical results and  competing theoretical models. 
Our results confirm that the shear driven jamming transition in Bagnoldian systems is well described by a critical scaling theory, and we relate this scaling theory to the phenomenological constituent laws for dilatancy and friction.

\end{abstract}
\pacs{83.80.Fg, 64.60.Ej, 45.70.-n}
\maketitle

\section{Introduction}
\label{secIntro}

The behavior of athermal ($T=0$) granular particles undergoing uniform shear flow has been much studied in different contexts, including both hard dry granular materials and soft materials such as foams, emulsions, and non-Brownian suspensions \cite{Lemaitre}.  For such shear driven systems, the control parameters may be viewed as the particle packing fraction $\phi$ and the shear strain rate $\dot\gamma$.
At sufficiently low strain rates $\dot\gamma$, at densities $\phi$ below jamming, such systems are generally found to have either a Newtonian rheology, with pressure $p$ and shear stress $\sigma$ proportional to $\dot\gamma$, or a Bagnoldian rheology \cite{Bagnold}, with $p,\sigma\propto\dot\gamma^2$.  It has been argued recently \cite{OlssonTeitelRheology} that it is the particular mechanism of energy dissipation in the system which determines which of these two rheologies a given system will display.

For a system with Newtonian rheology we can define the viscous transport coefficients, $p/\dot\gamma\equiv\eta_p$, $\sigma/\dot\gamma\equiv\eta_\sigma$.  For a system with Bagnoldian rheology  we can define the Bagnold transport coefficients, $p/\dot\gamma^2\equiv B_p$, $\sigma/\dot\gamma^2\equiv B_\sigma$.  These transport coefficients characterize the global rheological response to shearing.  In the limit of sufficiently small $\dot\gamma$ below jamming, these transport coefficients by definition become independent of $\dot\gamma$ \cite{OlssonTeitelRheology,Lemaitre,Lois,daCruz}, and hence depend only on the particle packing fraction $\phi$.  We will refer to this limit of sufficiently small $\dot\gamma$ as the ``hard-core" limit.


Upon increasing the packing fraction  $\phi$ to a critical value $\phi_J$, such granular systems undergo a shear-driven {\em jamming} transition \cite{Durian,OlssonTeitel2007,OlssonTeitelScaling} from a liquid to a rigid-but-disordered solid state.  In the hard-core limit, this transition is characterized by a divergence of the transport coefficients $p/\dot\gamma^n$ and $\sigma/\dot\gamma^n$ ($n=1$ for Newtonian, $n=2$ for Bagnoldian).  For soft-core particles above the jamming transition, these transport coefficients diverge as $\dot\gamma\to 0$, reflecting the existence of a finite yield stress in the solid state.

For {\em frictionless} particles, the jamming transition is generally believed to be continuous.  In the hard-core limit, transport coefficients diverge as a power-law of the distance from jamming, $p/\dot\gamma^n,\sigma/\dot\gamma^n\sim (\phi_J-\phi)^{-\beta}$, as $\phi$ increases to $\phi_J$ from below.   
For soft-core particles a critical scaling theory, in analogy with phase transitions in equilibrium systems, has been used \cite{OlssonTeitel2007,OlssonTeitelScaling,OH1,OH2,Hatano1,Hatano2,OH3} to give a unified description of the critical behavior of rheology as a function of both $\phi$ and $\dot\gamma$ in the neighborhood of the jamming transition.

The goal of the present work is to numerically simulate a simple granular model that displays Bagnoldian rheology ($n=2$), and carry out a scaling analysis of the resulting $p$ and $\sigma$ to determine the critical exponent $\beta$, and related critical parameters.  We emphasis that 
when we refer to the {\em critical exponent} $\beta$, we mean the exponent that characterizes the true algebraic divergence {\em asymptotically} close to the athermal jamming critical point, i.e. $T=0$, $\phi\to\phi_J$, $\dot\gamma\to 0$.  While this asymptotic region may be small (and indeed the present work argues it is), it is of fundamental interest because analogy with equilibrium critical phenomena leads one to expect that this asymptotic  exponent $\beta$ is {\em universal}, i.e. independent of microscopic details \cite{OlssonTeitelHB,VOT}.  
Determining the numerical value of $\beta$ then allows one to test competing theoretical models which make specific predictions for behavior in this asymptotic region about the critical point.

To determine the critical exponent $\beta$, it will be necessary to test that the data used in the analysis is indeed in the asymptotic critical region.
If one fits to numerical or experimental data that lies outside this true asymptotic critical region, one is liable to find  only {\em effective} values of the exponent that may vary depending on the range of data considered (as will be shown in Figs.~\ref{fitparams} and \ref{fitparams2}), or may depend on other microscopic details.  
Thus, in determining $\beta$ from data fitting or scaling collapses, it is essential to check the self consistency of the resulting value of $\beta$ (and other fitting parameters) by varying the window of data used in the fit, shrinking it ever closer to the critical point to see if parameter values  are systematically changing or if they remain stable.
Without such a test, the value of $\beta$ obtained from such an analysis is likely to be unreliable, even though the fit may seem very good (as will be illustrated in our Fig.~\ref{p-s-scaled}).  Very few of the prior works in the literature carry out such a test. Here we will show that, although we go to quite low shear strain rates $\dot\gamma$, comparable or smaller than other prior works, we cannot get close enough to the critical point so that a leading scaling analysis gives self-consistent results; rather it becomes necessary to include the next leading {\em correction-to-scaling} term to arrive at consistent numerical values of the critical parameters, as we have earlier found for systems with Newtonian rheology \cite{OlssonTeitelScaling}.

Although we expect that $\beta$ should be universal for a given class of rheology, we do not expect $\beta$ to be the same for Newtonian systems ($n=1$) as for Bagnoldian systems ($n=2$), and prior works \cite{OlssonTeitelScaling,OH1,OH2,Hatano1,Hatano2,OH3,OlssonTeitelHB,VOT,Andreotti,Lerner,Kawasaki,DeGiuli} are consistent with that.  For systems with Newtonian rheology, numerical works using a simple scaling analysis gave values of $\beta\approx1.65$, $2.2$ and $2.17$ in two dimensions \cite{OlssonTeitel2007,Andreotti,Lerner}, and $2.63$ in three dimensions \cite{Lerner}.
However other works, either going closer to the critical point or including corrections-to-scaling, found generally somewhat larger values, $\beta\approx2.77$, $2.58$ and $2.5$ in two dimensions
\cite{OlssonTeitelScaling,OlssonTeitelHB,VOT}, and $2.55$ in three dimensions \cite{Kawasaki}.  Recent theoretical work has predicted $\beta\approx2.83$ \cite{DeGiuli}.
The value of the exponent $\beta$, being a property of the hard-core limit, has been shown to be independent of the details of the elastic repulsive interaction between particles \cite{OlssonTeitelHB}, and independent of the mechanism of energy dissipation \cite{VOT}, 
provided the rheology remains Newtonian.  
Further discussion of the numerical value of the exponent $\beta$ for Newtonian systems, and its relation to earlier works, may be found in Ref.~\cite{Kawasaki}.

For systems with Bagnoldian rheology ($n=2$), the value of the corresponding exponent $\beta$ remains in dispute.  Otsuki and Hayakawa developed \cite{OH1,OH2} a phenomenological mean-field theory of the jamming transition that predicted the value $\beta=4$.  Numerical simulations \cite{OH1,OH2,Hatano1,Hatano2,OH3}, carried out by varying $\phi$ in a constant volume ensemble, have reported values of $\beta$ somewhat smaller than $4$, but seem perhaps to be approaching this prediction as the window of data analyzed shrinks closer to the jamming critical point.
However, simulations by Peyneau and Roux \cite{Roux}, using an ensemble at constant normal pressure, found significantly different results, equivalent to a value of $\beta\approx5$.  Recent theoretical work by DeGuili et al. \cite{DeGiuli} has argued for a value $\beta\approx 5.7$.

In the present work, we carry out a careful scaling analysis of the critical behavior of the Bagnold coefficients $B_p$ and $B_\sigma$, so as to try to resolve this discrepancy.  We use the same model of massive frictionless disks as in earlier studies \cite{OH1,OH2,Hatano1,Hatano2,OH3}, with a dissipation proportional to the normal component of the velocity difference between particles in contact, such as is known to result in Bagnoldian rheology.  Our simulations are carried out varying $\phi$ and $\dot\gamma$, shearing the system at constant volume.  
When we include the leading {\em correction-to-scaling} term in our analysis, we find that our results are consistent with those of Peyneau and Roux \cite{Roux}, and thus closer to the theoretical prediction of DeGiuli et al. \cite{DeGiuli} than to that of Ostuki and Hayakawa \cite{OH1,OH2}.

The remainder of this paper is organized as follows.  In Sec.~\ref{sModel} we present the details of our numerical model and simulations.  In Sec.~\ref{sScaling} we review the scaling ansatz for the shear-driven jamming transition, making a connection to the ``constituent equations" formulation common in the granular rheology community.  We discuss the scaling functions and corrections-to-scaling.  In Sec.~\ref{sPrevious} we review previous theoretical and numerical results for the exponent $\beta$ for Bagnoldian rheology.  In Sec.~\ref{sResults} we present our results and scaling analysis.  In Sec.~\ref{sConclusion} we summarize and present our conclusions.

\section{Model and Simulation Method}
\label{sModel}

We use a well studied model \cite{OHern} of frictionless, bidisperse, soft-core circular disks in two dimensions, with equal numbers of big and small particles with diameter ratio $d_b/d_s=1.4$.  Particles interact only when they come into contact,  in which case they repel with an elastic potential,
\begin{equation}
{\cal V}_{ij}(r_{ij})=\left\{
\begin{array}{cc}
\frac{1}{\alpha} k_e\left(1-r_{ij}/d_{ij}\right)^\alpha,&r_{ij}<d_{ij}\\
0,&r_{ij}\ge d_{ij}.
\end{array}
\right.
\label{eInteraction}
\end{equation}
Here $r_{ij}\equiv |\mathbf{r}_{ij}|$, where $\mathbf{r}_{ij}\equiv \mathbf{r}_i-\mathbf{r}_j$ is the center to center displacement from particle $j$ at position $\mathbf{r}_j$ to particle $i$ at $\mathbf{r}_i$, and $d_{ij}\equiv (d_i+d_j)/2$ is the average of their diameters.  In this work we will use the value $\alpha=2$, corresponding to a harmonic repulsion.
We will measure energy in units such that $k_e=1$.
The resulting elastic force on particle $i$ from particle $j$ is,
\begin{equation}
\mathbf{f}_{ij}^\mathrm{el}=-\dfrac{d{\cal V}_{ij}(r_{ij})}{d\mathbf{r}_i} =\frac{k_e}{d_{ij}} \left(1-\frac{r_{ij}}{d_{ij}}\right)^{\alpha-1} \mathbf{\hat r}_{ij},
\end{equation}
where $\mathbf{\hat r}_{ij}\equiv \mathbf{r}_{ij}/r_{ij}$ is the inward pointing normal direction at the surface of particle $i$

Particles also experience a dissipative force when they come into contact.  We take this force to be proportional to the projection of the velocity difference of the contacting particles onto the direction normal to the surface at the point of contact.  The dissipative force on particle $i$ from particle $j$ is, 
\begin{equation}
\mathbf{f}_{ij}^\mathrm{dis}=-k_d [(\mathbf{v}_i-\mathbf{v}_j)\cdot\mathbf{\hat r}_{ij}]\mathbf{\hat r}_{ij},
\label{efdis}
\end{equation}
where $\mathbf{v}_i\equiv d\mathbf{r}_i/dt$ is the velocity of particle $i$.  We have earlier \cite{OlssonTeitelRheology} denoted this model of dissipation by CD$_{n}$ for ``normal contact dissipation."  This dissipative force is well known to result in Bagnoldian rheology \cite{OlssonTeitelRheology, daCruz,OH1,OH2,Hatano1,Hatano2,OH3}.

Particle motion is governed by Newton's equation,
\begin{equation}
m_i\dfrac{d^2\mathbf{r}_i}{dt^2}=\sum_j\left[\mathbf{f}_{ij}^\mathrm{el}+\mathbf{f}_{ij}^\mathrm{dis}\right],
\label{eEqMotion}
\end{equation}
where $m_i$ is the mass of particle $i$ and the sum is over all particles $j$ in contact with particle $i$.  In this work we take particles to have a mass proportional to their area, i.e. small particles have mass $m_s$ and big particles mass $m_b$, with $m_b/m_s=(d_b/d_s)^2$.  

The above model possesses two important time scales \cite{OlssonTeitelRheology}, the elastic and dissipative relaxation times,
\begin{equation}
\tau_e\equiv\sqrt{m_sd_s^2/k_e},\qquad \tau_d\equiv m_s/k_d.
\end{equation}
The parameter
\begin{equation}
Q\equiv \tau_d/\tau_e=\sqrt{m_sk_e/(k_dd_s)^2}
\end{equation}
measures the elasticity of collisions; a head-on collision of two small particles will be totally inelastic (coefficient of restitution $e=0$) when $Q<1/2$.  In the present work we will measure  distance in units such that $d_s=1$, and time in units such that $\tau_e=1$ (hence, in these units, $m_s=1$).  Our simulations are in the strongly inelastic limit with $Q=1$, though the critical behavior sufficiently close to $\phi_J$ is expected \cite{OHL, VOT-CDn2} to be independent of the value of $Q$.

We simulate $N= 262144$ total particles in a box of fixed area $L^2$, using periodic Lees-Edwards boundary conditions \cite{LeesEdwards} to impose a uniform shear strain $\gamma(t)=\dot\gamma t$ with flow in the $x$-direction.  The box length $L$ is chosen to set the particle packing fraction,
\begin{equation}
\phi=\dfrac{\pi N}{2L^2}\left[\left(\dfrac{d_s}{2}\right)^2+\left(\dfrac{d_b}{2}\right)^2\right].
\end{equation}
Our system size is sufficient large that finite size effects are negligible for the range of parameters we consider, as we demonstrate explicitly in Appendix A.

To determine the global rheology of the system we measure the pressure tensor of each configuration.  We consider only the part arising from the elastic contact forces, since at the low strain rates $\dot\gamma$ considered here the elastic part dominates over the kinetic and dissipative parts.  The elastic contribution to the pressure tensor is \cite{OHern},
\begin{equation}
\mathbf{p}^\mathrm{el}\equiv L^{-2}\sum_{i<j}\mathbf{f}^\mathrm{el}_{ij}\otimes\mathbf{r}_{ij}.
\end{equation}
The average pressure and shear stress in the system are then,
\begin{equation}
p=\frac{1}{2}\left[\langle p^\mathrm{el}_{xx}\rangle +\langle p^\mathrm{el}_{yy}\rangle\right],\quad
\sigma = -\langle p^\mathrm{el}_{xy}\rangle.
\end{equation}
Here and in the following $\langle\dots\rangle$ represents an ensemble average over configurations in the steady state.

We integrate the equations of motion (\ref{eEqMotion}) using a modified velocity-Verlet algorithm with a Heun-like prestep to account for the velocity dependent acceleration. We use an integration time step of $\Delta t=0.1\tau_e$. We simulate over a range of strain rates from $\dot\gamma=10^{-4}$ down to $2\times 10^{-8}$, for a window of $\phi$ no greater than $1\%$ above and below $\phi_J$.  We shear to a total strain $\gamma$ that depends on the strain rate:  for $\dot\gamma \ge 10^{-5}$ we use $\gamma\sim 10 - 30$; for $\dot\gamma= 10^{-6}$ we use $\gamma\sim 2 - 10$; for $\dot\gamma= 10^{-7}$ we use $\gamma\sim 0.5 - 2$; for $\dot\gamma= 2\times 10^{-8}$ we use $\gamma\sim 0.5 - 0.8$, with the runs being longer the closer $\phi$ is to $\phi_J$.  Simulations at our largest $\dot\gamma$ are started from an initial random configuration at each $\phi$; simulations at smaller $\dot\gamma$ start from the ending configuration of the simulation at the next larger $\dot\gamma$, at the same value of $\phi$.
In each case we exclude the initial 20\% of the run in order to reach steady state, and then collect data for our averages from the remainder of the run.

\section{Critical Scaling}
\label{sScaling}

In this section we describe the theory of critical scaling that we will use to analyze our numerical data, discussing the scaling functions, critical exponents, and corrections-to-scaling.  We will also discuss the relation between this scaling theory and the empirical constituent equations that are often used to describe the rheology of hard-core particles.  Although our numerical simulations in the present work are for a system with Bagnoldian rheology, we frame the discussion here more generally, to deal with both Newtonian and Bagnoldian systems.

\subsection{The scaling ansatz}

The scaling ansatz \cite{OlssonTeitel2007,OlssonTeitelScaling} for describing critical behavior in the neighborhood of a continuous jamming transition is motivated by analogy with the renormalization group theory of equilibrium phase transitions.  It posits that, as one approaches close to the critical jamming point, there is a diverging length scale $\xi$ and that (i) the behavior of the system at different locations in the control parameter space is, to leading order, the same at equal values of $\xi$, and (ii) if one changes the control parameters so as to change the length scale $\xi$ by a factor $b$, $\xi^\prime=\xi/b$, all critical observables and control parameters will scale with the distance from their values at the critical point as some power of $b$; these powers define the critical exponents. As a consequence, critical observables are homogenous functions of the distance of the control parameters to the critical point.

For our simulations the control parameters are the packing fraction $\phi$ and the shear strain rate $\dot\gamma$.  The jamming transition is at $\phi=\phi_J$, $\dot\gamma=0$.  Our scaling variables are therefore $\delta\phi\equiv\phi-\phi_J$ and $\dot\gamma$.  Taking pressure as an example of an observable that displays critical behavior at the jamming transition, we can then write,
\begin{equation}
pb^{y/\nu}=f(\delta\phi b^{1/\nu}, \dot\gamma b^z, w_1 b^{-\omega_1}, w_2 b^{-\omega_2},\dots).
\label{escale0}
\end{equation}
In the above, the $w_i$ represent additional parameters that might describe other microscopic aspects of the system, for example a parameter controlling the dispersity of the particles.  We choose them such that at the critical point, $w_i=0$.  By assumption, the scaling function at the critical point is a constant.  

The parameters $\delta\phi$ and $\dot\gamma$ are said to be {\em relevant} variables; it is necessary to tune them to specific values, i.e. $\delta\phi=\dot\gamma=0$, to see the singular critical behavior.  The scaling exponents of relevant variables, in this case $1/\nu$ and $z$, are positive.  The parameters $w_i$ are said to be {\em irrelevant}; there is no need to tune them to any specific values to see the singular behavior.  The scaling exponents of irrelevant variables, in this case the $-\omega_i$, are negative (and so the $\omega_i$ are by definition positive). The {\em leading irrelevant variable} is the irrelevant variable whose scaling exponent has the smallest absolute value.  In our discussion below we will consider only the leading irrelevant variable.

To see how Eq.~(\ref{escale0}) leads to critical scaling, we now choose for the arbitrary scaling factor $b$ the specific value $b=\dot\gamma^{-1/z}$.  This gives,
\begin{equation}
p=\dot\gamma^{y/z\nu}f\left(\dfrac{\delta\phi}{\dot\gamma^{1/z\nu}}, 1, w\dot\gamma^{\omega/z}\right).
\label{escale1}
\end{equation}
Note that as the control parameters are tuned to the jamming transition, and so $\dot\gamma\to 0$, the dependence of $p$ on the variable $w$ vanishes as a consequence of the exponent $\omega/z>0$ (which follows since $w$ has a negative scaling exponent, $-\omega<0$).  This is why $w$ is called {\em irrelevant}, and why it is not necessary to explicitly tune the system to the value $w=0$ in order to explore the singular critical behavior.

Exactly at the jamming density, $\delta\phi=0$, the above gives as $\dot\gamma\to 0$ the non-linear rheology,
\begin{equation}
p=\dot\gamma^qf(0,1,0),\quad q\equiv y/z\nu,\quad \mathrm{at}\>\>\phi=\phi_J.
\label{eq1}
\end{equation}
Above the jamming density, where $\delta\phi>0$, we expect that $\lim_{\dot\gamma\to 0} p$ is just the finite yield stress $p_0$.  For Eq.~(\ref{escale1}) to be finite and independent of $\dot\gamma$ in this limit requires,
\begin{equation}
f(x,1,0)\sim x^y\quad\mathrm{as}\quad x\to +\infty,
\end{equation}
and gives,
\begin{equation}
p_0(\phi)=\lim_{\dot\gamma\to 0} p(\phi,\dot\gamma)\sim \delta\phi^y,\quad\mathrm{for}\>\>\phi>\phi_J.
\label{ep01}
\end{equation}
Hence the exponent $y$  determines how the yield stress vanishes as $\phi$ decreases to $\phi_J$ from above.

Below the jamming density, where $\delta\phi<0$, we expect that $\lim_{\dot\gamma\to 0}p\sim\dot\gamma^n$, where $n=1$ for Newtonian rheology and $n=2$ for Bagnoldian. For Eq.~(\ref{escale1}) to agree with this behavior then requires,
\begin{equation}
f(x,1,0)\sim |x|^{-(z\nu n-y)}    \quad\mathrm{as}\quad x\to -\infty,
\end{equation}
and gives,
\begin{equation}
\lim_{\dot\gamma\to 0} p/\dot\gamma^n\sim |\delta\phi|^{-\beta},\quad\beta\equiv z\nu n-y,\quad\mathrm{for}\>\>\phi<\phi_J,
\label{ebeta1}
\end{equation}
where the exponent $\beta$ gives the divergence of the hard-core transport coefficient as $\phi$ increases to $\phi_J$ from below.  

Note that in Eqs.~(\ref{eq1}) and (\ref{ebeta1}) for the exponents $q$ and $\beta$, the exponents $\nu$ and $z$ enter only in the combination $z\nu$.  Thus the non-linear rheology at $\phi_J$ (given by $q$), the vanishing of the yield stress above $\phi_J$ (given by $y$), and the divergence of the transport coefficient below $\phi_J$ (given by $\beta$), are all determined by just two exponent combinations, $y$ and $z\nu$.

Since the exponents $y$ and $q$ are determined by behavior above and exactly at $\phi_J$, where the softness of the particles is an essential feature (strictly hard-core particles cannot be compressed above $\phi_J$, nor sheared at a finite rate $\dot\gamma$ exactly at $\phi_J$), it is expected that $y$ and $q$ will depend on details of the soft-core interaction potential \cite{OHern}, and hence the exponent $\alpha$ in Eq.~(\ref{eInteraction}). However, since the exponent $\beta$ is determined from behavior in the hard-core limit below $\phi_J$, we expect that $\beta$ will not depend on the interaction exponent $\alpha$; we have explicitly verified this in simulations of a Newtonian system \cite{OlssonTeitelHB}.

Finally, we note that if one is sufficiently close to the jamming point, so that $w\dot\gamma^{\omega/z}$ is small enough to be ignored, then Eq.~(\ref{escale1}) predicts that data at different values of $\phi$ and $\dot\gamma$ will all collapse to a single curve if plotted as,
\begin{equation}
\dfrac{p}{\dot\gamma^{y/z\nu}}\quad\mathrm{vs}\quad \dfrac{\delta\phi }{\dot\gamma^{1/z\nu}}.
\label{ecollapse}
\end{equation}
Such a collapse is the defining signature of the critical scaling theory; a single scaling function $f(x,1,0)$ unites behavior above, below, and at the transition $\phi_J$, as a function of both control variables $\phi$ and $\dot\gamma$.  Testing for such a collapse provides one way to numerically determine the exponent combinations $q=y/z\nu$ and $1/z\nu$, and hence $y=qz\nu$ and $\beta=z\nu n-y=(n-q)z\nu$.

Another key assertion of the critical scaling theory is that the exponents $\nu$, $z$ and $\omega$ have the same values, independent of which observable is being measured.  $\nu$ is known as the correlation length critical exponent, $z$ the dynamic critical exponent, and $\omega$ the correction-to-scaling critical exponent.  The exponent $y$ is specific to the observable being measured.  In this work we will be concerned with the scaling of the pressure $p$ and the shear stress $\sigma$.  It is generally assumed that, since $p$ and $\sigma$ are both components of a unified tensor, their scaling exponents $y$ are the same.  This has been confirmed numerically for the case of Newtonian rheology \cite{OlssonTeitelScaling}, and we confirm in the present work that this is also the case for Bagnoldian rheology.

\subsection{Corrections-to-scaling}

The scaling collapse of Eq.~(\ref{ecollapse}) will only hold if $w\dot\gamma^{\omega/z}$ is small enough to be ignored; this will always be true sufficiently close to the jamming transition.  However, since $w$ is not directly tuned in the simulation (and indeed it may not even be known what physical features of the system are represented by the parameter $w$) it may be that this term is not sufficient small over much of the range of control parameters $\phi$ and $\dot\gamma$ where simulations are feasible.   In this case one must take into account the finite effects of the leading irrelevant variable, and these are known as {\em corrections-to-scaling} \cite{Binder,Hasenbusch}.  Corrections-to-scaling have been found to be important in equilibrium spin-glass problems \cite{Hasenbusch}, and we have previously shown them to be important for Newtonian rheology near jamming \cite{OlssonTeitelScaling}.

In this case one can expand Eq.~(\ref{escale1}) about $w=0$ for small but finite $w$ to get,
\begin{equation}
p=\dot\gamma^{y/z\nu}\left[f_1\left(\dfrac{\delta\phi}{\dot\gamma^{1/z\nu}}\right)+
\dot\gamma^{\omega/z}f_2\left(\dfrac{\delta\phi}{\dot\gamma^{1/z\nu}}\right)\right].
\label{ecorrection}
\end{equation}
The first term is the leading scaling term and gives the results discussed in the previous section.  The second term is the leading correction-to-scaling term, and $\omega$ is the correction-to-scaling exponent.
Because of the prefactor in front of the second scaling term $f_2$, a simple data collapse as in Eq.~(\ref{ecollapse}) will no longer hold, and one must fit data to the above more complicated form in order to determine the critical exponents.  

The correction-to-scaling term effects the three limiting critical behaviors of the previous section as follows.  Exactly at $\phi_J$, where $\delta\phi=0$, Eq.~(\ref{ecorrection}) becomes,
\begin{equation}
p=\dot\gamma^q\left[f_1(0)+\dot\gamma^{\omega/z}f_2(0)\right],\quad q\equiv y/z\nu,
\label{eq2}
\end{equation}
giving a correction to Eq.~(\ref{eq1}) for the asymptotic power-law relation for the rheology   as $\dot\gamma$ increases at $\phi=\phi_J$.

For the limiting behaviors as $\dot\gamma\to 0$ above and below $\phi_J$, it is easiest to return to Eq.~(\ref{escale0}) and choose $b=|\delta\phi|^{-\nu}$, to get,
\begin{equation}
p=|\delta\phi|^yf\left(\pm1,\dfrac{\dot\gamma}{|\delta\phi|^{z\nu}},w|\delta\phi|^{\omega\nu}\right).
\end{equation}
Expanding in $w$ then gives,
\begin{equation}
p=|\delta\phi|^y\left[\tilde f_{1\pm}\left(\dfrac{\dot\gamma}{|\delta\phi|^{z\nu}}\right)+
|\delta\phi|^{\omega\nu}\tilde f_{2\pm}\left(\dfrac{\dot\gamma}{|\delta\phi|^{z\nu}}\right)\right],
\label{epx}
\end{equation}
where $\pm$ denote above and below $\phi_J$ respectively.

For $\phi>\phi_J$, such that $\delta\phi>0$, we expect $p$ to approach the finite yield stress $p_0$ as $\dot\gamma\to 0$, hence we expect $\tilde f_{1+}(0)$ and $\tilde f_{2+}(0)$ to be finite, and so,
\begin{equation}
p_0(\phi)=\delta\phi^y\left[\tilde f_{1+}(0)+\delta\phi^{\omega\nu}\tilde f_{2+}(0)\right],
\label{ep02}
\end{equation}
giving a correction to Eq.~(\ref{ep01}) for the vanishing of the yield stress as $\phi\to\phi_J$ from above.

For $\phi<\phi_J$, such that $\delta\phi<0$, we expect $p\sim\dot\gamma^n$ as $\dot\gamma\to 0$, hence we expect $\tilde f_{1-}(x)\sim\tilde f_{2-}(x)\sim x^n$ as $x\to 0$, and so,
\begin{equation}
p/\dot\gamma^n=|\delta\phi|^{-\beta}\left[C_{p1}+|\delta\phi|^{\omega\nu}C_{p2}\right],\quad\beta\equiv z\nu n-y,
\label{ebeta2}
\end{equation}
with $C_{p1}$ and $C_{p2}$ constants,
giving a correction to Eq.~(\ref{ebeta1}) for the divergence of the hard-core transport coefficient as $\phi\to\phi_J$ from below.  Note that in all three cases, the correction term is governed by the same correction-to-scaling exponent $\omega$, and that the relative contribution of the correction term vanishes as $\phi_J$ is approached, i.e. as $\delta\phi\to 0$.

We have earlier shown how corrections-to-scaling are crucial for a consistent understanding of the behavior of systems with Newtonian rheology \cite{OlssonTeitelScaling}.  Independent simulations by Kawasaki et al. \cite{Kawasaki} have recently confirmed this.  In the present work we will show that it is also necessary to consider corrections-to-scaling for systems with Bagnoldian rheology, for the parameter range that is typically simulated.

A final note on the preceding scaling theory is in order: In equilibrium phenomena, another form of scaling corrections may occur in the special case when the system dimension $d$ is exactly equal to the {\em upper critical dimension} $d_{uc}$.  For $d<d_{uc}$, fluctuations are important and dimensionality can affect the value of critical exponents.  For $d>d_{uc}$ fluctuations are unimportant, {\em mean-field} results describe  the transition well, and critical exponents become independent of the dimension $d$.  When $d=d_{uc}$, logarithmic corrections are believed to modify the scaling variables.  As it has been suggested that $d_{uc}=2$ for the jamming transition, we discuss this possibility of logarithmic corrections in Appendix B.

\subsection{The constituent equations}
\label{sConstituent}

The above scaling approach has been framed in terms of the packing fraction $\phi$ and strain rate $\dot\gamma$, which are the control parameters of our, and many earlier, simulations of soft-core particles.  For the rheology of hard-core particles, however, studies are often done at constant pressure rather than constant volume, and it has been common to introduce as the control parameter the quantity,
\begin{equation}
I\propto \dot\gamma/p^{1/n}.
\label{eInertialNo}
\end{equation}
For Bagnoldian rheology with $n=2$, $I$ is referred to as the {\em inertial number}  \cite{Forterre}.  For Newtonian rheology with $n=1$, $I$ is referred to as the {\em viscous number} \cite{Boyer}.  Because $I$ is defined for the hard-core limit below the jamming transition, where the pressure obeys the strict relation $p\propto \dot\gamma^n$, 
the transport coefficient $p/\dot\gamma^n\propto1/I^n$ is independent of $\dot\gamma$ and only varies with the packing fraction $\phi$.  Hence there is a unique mapping between $I$ and $\phi$ and so the behavior of the system depends only on the value of $I$ and not the specific values of $p$ and $\dot\gamma$ separately. Further, $I=0$ locates the jamming transition.

The rheology in this hard-core limit {\em below} jamming is then characterized by two empirical {\em constituent equations}, which in the limit of small $I$ can be written as \cite{Forterre,Boyer},
\begin{align}
\phi_J-\phi(I)&\propto I^a,\label{ec1}
\\
\mu(I)-\mu_J&\propto I^b.\label{ec2}
\end{align}
Here $\phi(I)$ is the packing fraction at control parameter $I$, and $\mu(I)\equiv\sigma/p$ is the effective macroscopic friction of the system, which in general is finite even though the particles in our model are themselves frictionless; $\mu_J$ is the value of $\mu$ at the jamming transition.  The first of the two constituent equations is often referred to as the {\em dilatancy law}, while the second is the {\em friction law}.

We now show how these constituent equations may be derived from the critical scaling theory, and how the exponents $a$ and $b$ are related to the critical exponents $\nu$, $z$, $y$ and $\omega$.   Equation~(\ref{ec1}) follows directly from Eq.~(\ref{ebeta2}).  We have, to lowest order in the correction-to-scaling,
\begin{equation}
I \equiv \lim_{\dot\gamma\to 0}[(\dot\gamma^n/p)^{1/n}]=|\delta\phi|^{\beta/n}C_{p1}^{-1/n}[1-|\delta\phi|^{\omega\nu}C_{p2}/nC_{p1}].
\end{equation}
Inverting the above to write $|\delta\phi|$ in terms of $I$, we get to lowest order,
\begin{equation}
|\delta\phi|=\phi_J-\phi =I^{n/\beta}\left[C_{\phi 1}+C_{\phi 2}I^{\omega\nu n/\beta}\right].
\label{edphi}
\end{equation}
Thus Eq.~(\ref{ec1}) represents the leading term above as $I\to 0$, and
\begin{equation}
a=n/\beta.
\label{ea}
\end{equation}

To get the second constituent equation we just note that the shear stress scales similarly to the pressure in Eq.~(\ref{ebeta2}), i.e. as $\dot\gamma\to 0$, 
\begin{equation}
\sigma/\dot\gamma^n=|\delta\phi|^{-\beta}\left[C_{\sigma 1}+|\delta\phi|^{\omega\nu}C_{\sigma 2}\right],
\end{equation}
so we can write for the hard-core limit $\dot\gamma\to 0$,
\begin{equation}
\mu\equiv \dfrac{\sigma}{p}=\dfrac{C_{\sigma 1}+C_{\sigma 2}|\delta\phi|^{\omega\nu}}{C_{p 1}+C_{p 2}|\delta\phi|^{\omega\nu}}.
\end{equation}
Because $p$ and $\sigma$ both scale to leading order with the same exponent $y$ (and hence the same $\beta=z\nu n-y$), the variation of $\mu$ with $\phi$ is due entirely to the correction-to-scaling terms, depending on the correction-to-scaling exponent $\omega$.  Expanding the above to lowest order in $|\delta\phi|$ we get,
\begin{equation}
\mu = \mu_J + C_\mu|\delta\phi|^{\omega\nu}
\label{emuphi}
\end{equation}
where $\mu_J \equiv C_{\sigma 1}/C_{p1}$ is the value when $\delta\phi\to0^-$, i.e. as jamming is approached from below. Substituting in for $|\delta\phi|$ from Eq.~(\ref{edphi}) then gives,
\begin{equation}
\mu-\mu_J=I^{\omega\nu n/\beta}\left[C_{\mu 1}+C_{\mu 2}I^{\omega\nu n/\beta}\right].
\end{equation}
Thus Eq.~(\ref{ec2}) represents the leading term above as $I\to 0$, and,
\begin{equation}
b=\omega\nu n/\beta = \omega\nu a.
\label{eb}
\end{equation}

\section{Summary of Previous Results}
\label{sPrevious}

The critical exponents of the static jamming transition arising from compression or quenching have been found to be independent of the dimensionality of the system \cite{OHern}.  A similar result has been claimed numerically  \cite{OH1,OH2}, and argued theoretically \cite{DeGiuli}, for the shear driven jamming transition. In this section we therefore review prior results from both two and three dimensional simulations, although our own work reported here has been in two dimensions.

Numerous simulations have been carried out by others on the model of spherical particles interacting with the elastic and dissipative forces described in Sec.~\ref{sModel}.  
Many of these simulations are for particles that include tangential frictional forces in their interactions.  Here we focus on those simulations that are for frictionless particles, such as those we study in the present work.  
We first consider those simulations carried out in an ensemble at fixed volume, where $\phi$ and $\dot\gamma$ are the simulation control parameters.  We then consider simulations carried out in an ensemble at fixed pressure $p$.

\subsection{Constant volume simulations}
\label{sPrevVol}

Simulations by Garcia-Rojo et al. \cite{Rojo} suggested that, at low packing fractions, the shear viscosity $\sigma/\dot\gamma$ diverged as $1/(\phi_c-\phi)$, with $\phi_c<\phi_J$.  However, later work \cite{OH1,OH2} argued that this conclusion was an artifact of not probing closely enough to the jamming transition $\phi_J$; it was later shown that the true scaling region near $\phi_J$ shrinks in size as particles become increasingly elastic \cite{OHL}.

Hatano \cite{Hatano1} studied essentially the same bidisperse model as described in Sec.~\ref{sModel}, simulating in three dimensions for the case of harmonic ($\alpha=2$) and Hertzian ($\alpha=5/2$)  interactions, with elasticity parameters $Q=10$ and $Q=100$, respectively.  Using $N=1000$ particles and exploring a window of packing fraction $|\delta\phi|/\phi_J\approx 0.1$ and strain rate range $10^{-4}\le\dot\gamma\tau_e \le 1$, he collapsed his data according to a common scaling curve (similar to our Eq.~(\ref{ecollapse}), but using instead scaling variables $p/|\delta\phi|^y$ and $\dot\gamma/|\delta\phi|^{z\nu}$), and claimed evidence for exponents $y=1.2$, $\beta=2.6$  for $\sigma$ and $y=1.2$, $\beta=3.0$ for  $p$, for the harmonic interaction; and $y=1.8$, $\beta=3.0$ for both $\sigma$ and $p$ for the Hertzian interaction.

Otsuki and Hayakawa \cite{OH1,OH2} developed a phenomenological mean-field like theory for the exponents describing the rheology of Bagnoldian systems.  Defining $\Delta\equiv\alpha-1$ as the power law for the repulsive interaction force, they have predicted the exponents $y=\Delta$ and $\beta=4$, the latter being independent of $\Delta$.  For the harmonic interaction with $\Delta=1$, our Eqs.~(\ref{eq1}) and (\ref{ebeta1}) would then lead to the conclusion $1/z\nu = 2/(\beta+y)=2/5$, and $q=y/z\nu=2/5$.

To numerically test these predictions, Otsuki and Hayakawa \cite{OH1,OH2} then carried out numerical simulations of the same model as that used here, in the strongly inelastic limit with $Q=1$, considering several different examples of the size dispersity of particles, in two, three and four dimensions, for both harmonic ($\alpha=2$) and Hertzian ($\alpha=5/2$) interactions. They used systems with more particles and much slower strain rates than Hatano \cite{Hatano1}, with
up to $N=4000$ particles and $5\times 10^{-7}\le\dot\gamma\tau_e\le 5\times 10^{-5}$ in two dimensions.  They argued that their results agreed with their theoretical predictions, however they demonstrated this only by data collapses (using the same scaling variables as Hatano), in which they used the assumed values of the critical exponents and a predetermined estimate of $\phi_J$.  No independent data fitting to determine the best fitted values of the exponents and $\phi_J$ were performed.  Since the best fitted values of exponents can depend very sensitively on the value taken for $\phi_J$, the scaling collapses of Refs.~\cite{OH1} and \cite{OH2} cannot be taken as conclusive.

The mean-field theory of Otsuki and Hayakawa \cite{OH1,OH2} also involves as a key assumption that the relevant time scale for the rheology at a packing fraction $\phi$ is set by the frequency $\omega^*$ that marks the low frequency edge of the plateau (``boson peak") in the density states of elastic vibrations of the statically jammed solid at $\phi$ \cite{Wyart}.  This frequency scales as $\omega^*\sim\delta\phi^{\Delta/2}$ \cite{Liu+Nagel}.  However, Lerner et al. \cite{Lerner} have shown that, for a sheared system with Newtonian rheology, there is a unique isolated mode below $\omega^*$ that is responsible for the diverging time scale of the shearing rheology; thus it is reasonable to wonder if the same might be true for Bagnoldian rheology, and hence the relevant time scale  may behave differently from that assumed by Otsuki and Hayakawa \cite{OH1,OH2}.

More recently, Hatano has repeated his earlier simulations \cite{Hatano2} for the harmonic interaction in three dimensions, but now using $N=4000$ particles and a smaller window of packing fractions, $|\delta\phi|/\phi_J\approx 0.023$, and smaller range of strain rates, $10^{-7}\le\dot\gamma\tau_e \le 10^{-2}$. He then finds exponents $y=1.5$, $\beta=3.5$ for $\sigma$ and $y=1.5$, $\beta=3.9$ for $p$.
Otsuki and Hayakawa have similarly repeated their simulations \cite{OH3} for a polydisperse system of $N=4000$ particles with the harmonic interaction in two dimensions.  For a packing fraction window of $|\delta\phi|/\phi_J\approx 0.024$, and a strain rate range of $5\times 10^{-7}\le\dot\gamma\tau_e\le 5\times 10^{-5}$, they now fit their data to a scaling form with $\phi_J$ and exponents as free fitting parameters.  They then find $y=1.09$, $\beta=3.56$ from $\sigma$ and $y=1.06$, $\beta=3.59$ from $p$.  

Summarizing these previous simulations at constant volume, it appears that the value for the transport coefficient exponent $\beta$ is increasing as the data gets restricted to a smaller window about the critical jamming point (i.e. smaller $\dot\gamma$ and smaller $|\delta\phi|/\phi_J$).  Moreover, this value is {\em perhaps} approaching the Otsuki and Hayakawa mean-field prediction \cite{OH1,OH2} of $\beta=4$.

\subsection{Constant normal pressure simulations}

Simulations have also been carried out in an ensemble at constant normal pressure $p$, rather than constant volume.  By ``normal pressure" we mean the pressure on surfaces for which the unit normal direction is orthogonal to the direction of the shear flow.  These simulations use
particle stiffnesses $k_e$ and strain rates $\dot\gamma$ that are thought to put the system in the hard-core limit where the inertial number $I$ of Eq.~(\ref{eInertialNo}) is independent of $\dot\gamma$ and $p$ separately, but depends only on the ratio $I\sim \dot\gamma/\sqrt{p}$.  To put $I$ into dimensionless form, we follow convention \cite{Forterre} and for a Bagnoldian system use $I=\dot\gamma\sqrt{m/pd}$, with $m$ and $d$ the mass and diameter of a typical particle, respectively.
Measuring the ensemble averaged packing fraction $\langle\phi\rangle$ and macroscopic friction $\mu=\langle\sigma\rangle/p$ then determines the exponents $a$ and $b$ via  the constituent equations  (\ref{ec1}) and (\ref{ec2}).  

It is often argued \cite{Forterre,Pouliquen,Lemaitre} that the constituent equations (\ref{ec1}) and (\ref{ec2}) are linear in $I$ at small $I$ , i.e. $a=b=1$.  In terms of the discussion of Sec.~\ref{sConstituent} this would imply a transport coefficient exponent $\beta=2/a=2$, and $\omega\nu=1$.  However this claim is best supported by results for particles with microscopic {\em frictional} interactions, rather than the {\em frictionless} particles considered here. That the constituent equations for frictional and frictionless particles involve different exponents is nicely illustrated  in Ref.~\cite{Bouzid} for the macroscopic friction $\mu$.
 
Early simulations by da Cruz et al. \cite{daCruz} for a two dimensional polydisperse system with harmonic elastic interaction, considered both frictional and frictionless particles.  For systems with up to $N=5000$ particles and a range of inertial number, $6\times 10^{-4}\le I\le 0.3$, they claimed that the packing fraction $\phi$ remained a linear function of $I$ (hence $a=1$) for both frictional and frictionless cases.  For frictional particles, $\mu-\mu_J$ was found to be linear in $I$, but for frictionless particles it was claimed to be sublinear, though no exponent value for $b$ was given.

Hatano \cite{Hatano3} has carried out simulations in three dimensions with $N=10000$ polydisperse frictionless particles using both the harmonic and Hertzian interactions with elastic parameter $Q=1$.  Fitting to a range of  inertial number, $10^{-5}\le I\le 0.5$, he  finds the exponents $a\approx 0.56\pm0.02$ and $b\approx 0.28\pm0.05$, for both interactions.  By Eqs.~(\ref{ea}) and (\ref{eb}) these values translate into the transport coefficient exponent $\beta=2/a\approx 3.57\pm0.13$ and $\omega\nu=b/a\approx 0.5\pm0.1$.  

Similar simulations have been carried out by Peyneau and Roux \cite{Roux} with up to $N=4000$ strongly inelastic monodisperse particles in three dimensions using the Hertzian interaction.  Fitting to a range of inertial number, $10^{-5}\le I\le 10^{-2}$, they find exponents $a\approx 0.40\pm0.02$ and $b\approx 0.39\pm0.02$, giving $\beta\approx 5.0\pm0.3$ and $\omega\nu\approx 1.0\pm 0.1$.

Most recently, DeGuili et al. \cite{DeGiuli} have proposed theoretical arguments that the exponents $a$ and $b$ for the constituent equations are the same for Bagnoldian rheology as for Newtonian rheology.  If so, then since $\beta=n/a$, we expect $\beta_\mathrm{Bagnold} = 2\beta_\mathrm{Newton}$.  
Using scaling arguments based on the distribution of contact forces at the static jamming transition,
as found numerically in two and three dimensions \cite{Lerner2,DeGiuli2} and as computed exactly within an infinite dimensional mean-field calculation \cite{Charbonneau,Charbonneau2},
DeGuili et al. predict the value $a\approx 0.35$, thus giving for Bagnold rheology $\beta\approx 5.7$.  They also predict $a=b$, and so $\omega\nu=1$.

Because the works summarized in this subsection claim to be in the hard-core limit, they cannot give any information about the critical exponents $y$, $1/z\nu=2/(\beta+y)$, or $q=y/z\nu$, which describe behavior at or above jamming.  However, if one assumes the value $y=1$ for the harmonic interaction (as done by Otsuki and Hayakawa \cite{OH1,OH2}, and as is believed to be the case for static, compression-driven, jamming \cite{OHern}), then one can obtain values for $1/z\nu$ and $q$; using DeGuili et al.'s value of $\beta\approx 5.7$, we would have $1/z\nu=q\approx 0.3$.

\section{Results}
\label{sResults}

In this section we present our results for the pressure $p$ and shear stress $\sigma$, as functions of the packing fraction $\phi$ and shear strain rate $\dot\gamma$, using the model and simulation methods described in Sec.~\ref{sModel}.  Because we will be fitting our data to scaling expressions such as Eq.~(\ref{ecorrection}), for which we do not a priori know the detailed form of the scaling functions, we wish to do our simulations in the region of the parameter space where the scaling variable $x= \delta\phi/\dot\gamma^{1/z\nu}$ is small, so that we may use expansions of the scaling function at small $x$ to do the fitting.  Thus as we decrease $\dot\gamma$, we restrict data to a decreasing window of $\phi$ about the jamming $\phi_J$.   

\begin{figure}[h!]
\includegraphics[width=3.2in]{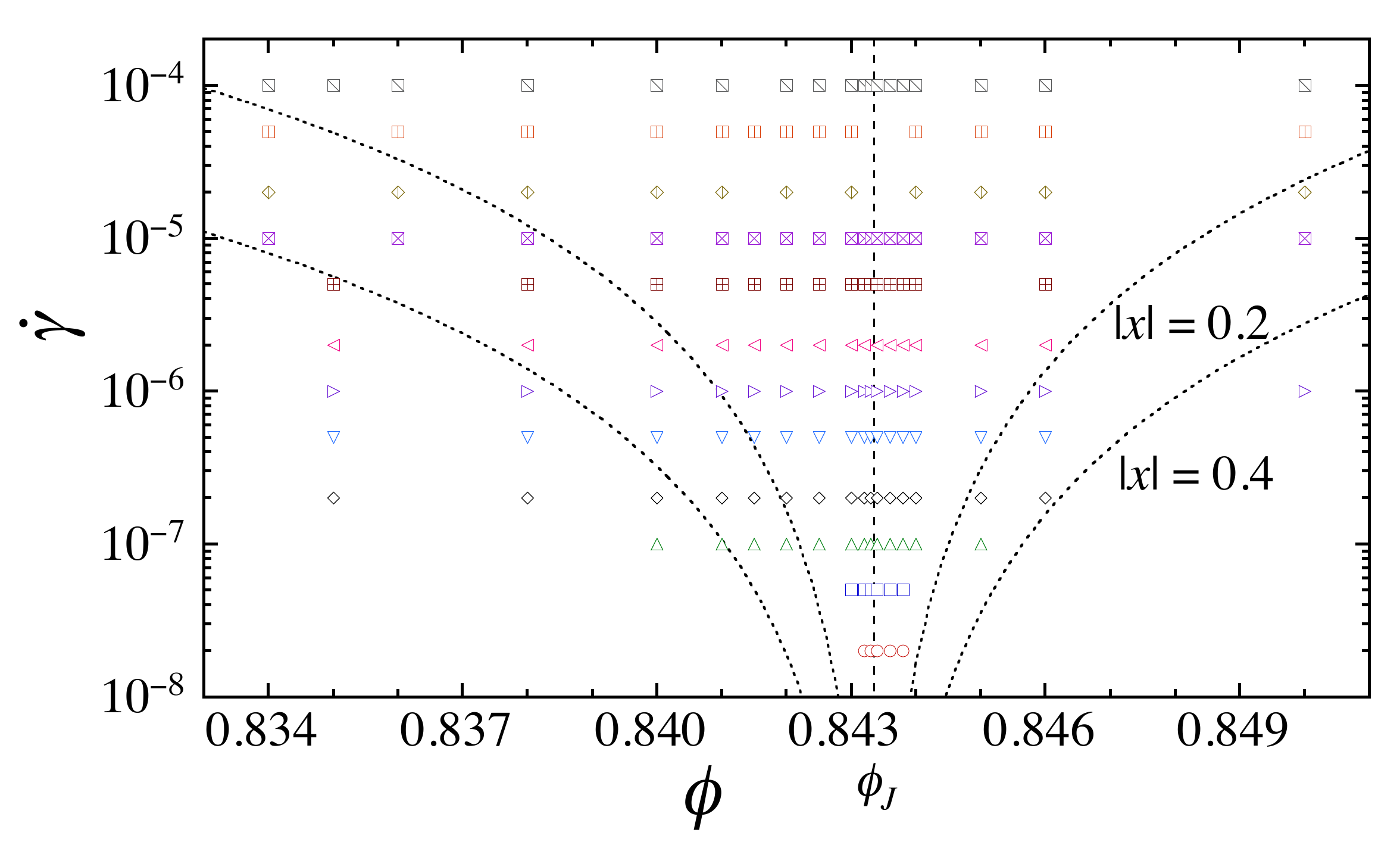}
\caption{(Color online) Control parameter phase space $(\phi,\dot\gamma)$.  Data points indicate the locations of control parameters used in our simulations.  Points with the same shape and color are at a common value of $\dot\gamma$.  
Curved dotted lines indicate contours of constant scaling variable $|x|=|\delta\phi|/\dot\gamma^{1/z\nu}= 0.2, 0.4$.
The vertical dashed line indicates the location of the jamming $\phi_J$ at $x=0$.
We have used the values $\phi_J=0.84335$ and $1/z\nu=0.32$ to define $x$.
}
\label{params}
\end{figure}

\begin{figure}[h!]
\includegraphics[width=3.2in]{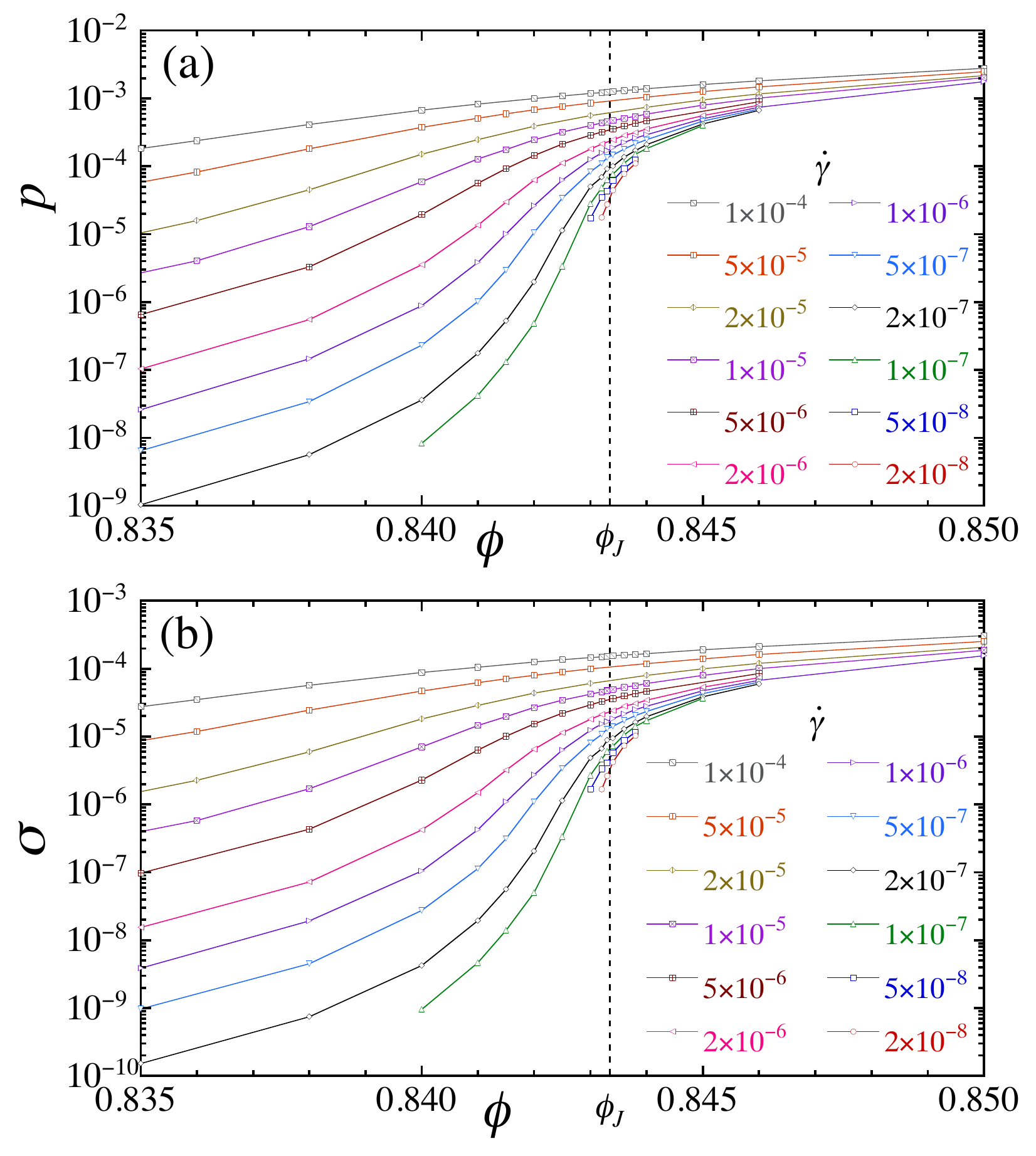}
\caption{(Color online) (a) Pressure $p$ and (b) shear stress $\sigma$ vs packing fraction $\phi$ at different values of the applied shear strain rate $\dot\gamma$.  The strain rate $\dot\gamma$ {\em decreases} as curves go from top to bottom.  The vertical dashed line indicates the location of the jamming $\phi_J$.
Error bars are smaller than the size of the data symbols, and are not shown.
}
\label{p-s-raw}
\end{figure}

\begin{figure}[h!]
\includegraphics[width=3.2in]{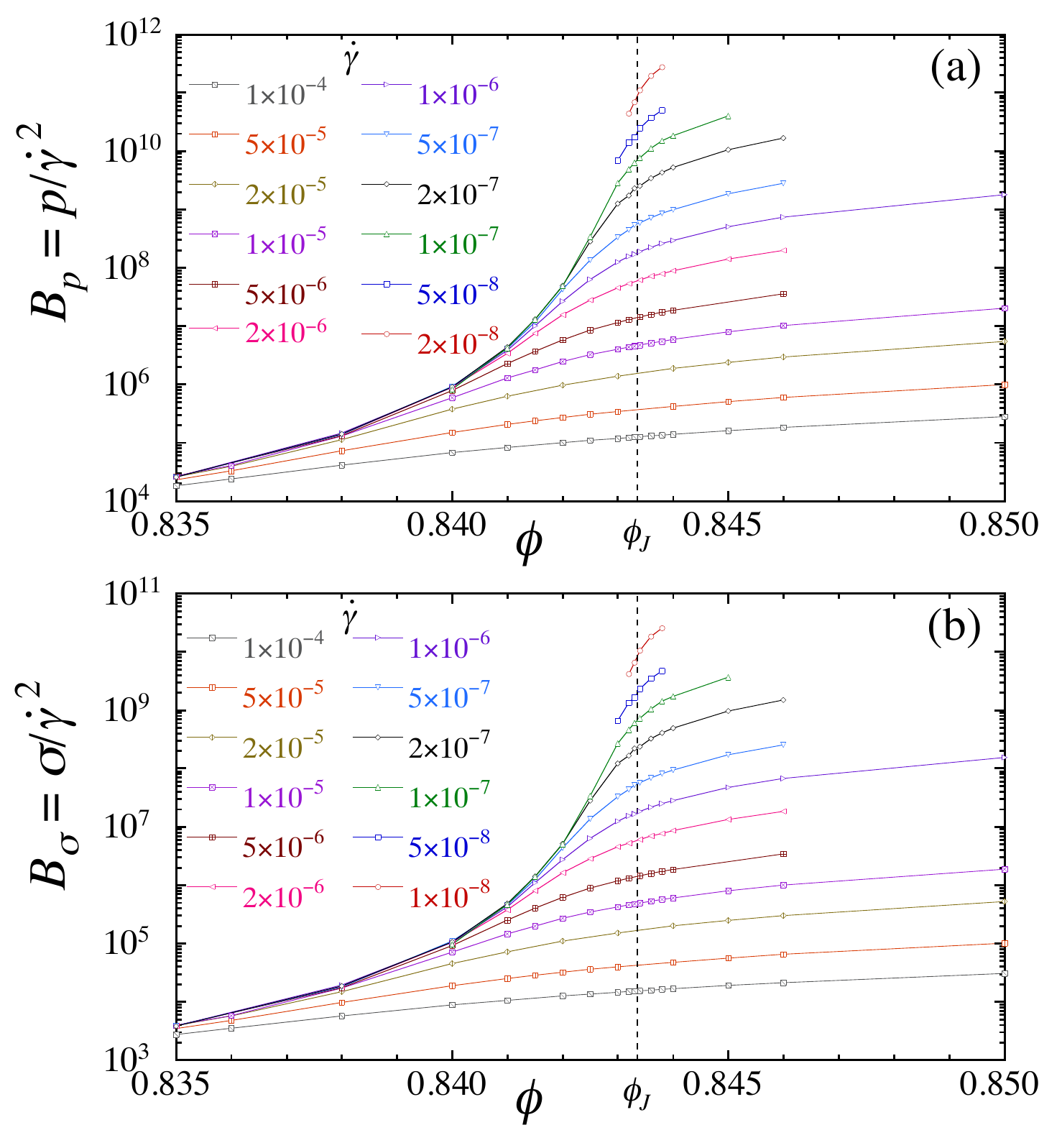}
\caption{(Color online) Bagnold coefficients for (a) pressure, $B_p\equiv p/\dot\gamma^2$, and (b) shear stress, $B_\sigma\equiv \sigma/\dot\gamma^2$ vs packing fraction $\phi$ at different values of the applied shear strain rate $\dot\gamma$.  The strain rate $\dot\gamma$ {\em increases} as curves go from top to bottom.  The vertical dashed line indicates the location of the jamming $\phi_J$.
Error bars are smaller than the size of the data symbols, and are not shown.
}
\label{Bp-Bs}
\end{figure}

We have considered in this work strain rates in the interval $2\times 10^{-8}\le \dot\gamma\le 10^{-4}$, going to lower rates than previous simulations.  In Fig.~\ref{params} we indicate the specific parameter points $(\phi,\dot\gamma)$ at which we have done our simulations; the colors and symbol shapes shown in this figure may be used to identify data points in subsequent plots.  The vertical dashed line in Fig.~\ref{params} (and in subsequent Figs.~\ref{p-s-raw} and \ref{Bp-Bs}) indicates the location of $\phi_J$.
The curved dotted lines represent contours of constant  scaling variable $|x|=|\delta\phi|/\dot\gamma^{1/z\nu}$.  We have used here the values $\phi_J= 0.84335$ and  $1/z\nu=0.32$, as determined by our analysis below.

In Fig.~\ref{p-s-raw} we plot our raw results for $p$ and $\sigma$ vs $\phi$, for different values of $\dot\gamma$.   Our data for $p$ and $\sigma$ span roughly six orders of magnitude.  In Fig.~\ref{Bp-Bs} we replot these data in terms of the Bagnold coefficients, $B_p\equiv p/\dot\gamma^2$ and $B_\sigma\equiv \sigma/\dot\gamma^2$.  The data at $\phi<\phi_J$ are seen to collapse to a common curve as $\dot\gamma$ decreases, confirming that our system does indeed have Bagnoldian rheology; this common curve as $\dot\gamma\to 0$ represents the hard-core limit.
As $\phi$ increases to $\phi_J$, the strain rate $\dot\gamma^*$ below which this hard-core limit is attained is seen to decrease; the scaling theory of the preceding section predicts $\dot\gamma^*\sim |\delta\phi|^{z\nu}$.  

For our units in which $m_s=d_s=1$, we have for the inertial number $I=1/\sqrt{B_p}$.  Noting the range in Fig.~\ref{Bp-Bs}a over which we have data in the hard-core limit, we see that our simulations allow us to probe a range of inertial numbers $5\times 10^{-5}<I<6\times 10^{-3}$, with our smallest value of $I$ somewhat larger than that used by Peyneau and Roux \cite{Roux}.  However an important virtue of the scaling function approach is that it unifies the hard-core behavior below $\phi_J$ with the soft-core behavior approaching and above $\phi_J$;  it thus lets us use  data outside the hard-core limit in order to determine the exponent $\beta$ that characterizes the hard-core rheology.

\subsection{Without corrections-to-scaling}
\label{ssWithout}

We will first attempt to fit our data to the scaling form ignoring corrections-to-scaling, i.e. to Eq.~(\ref{ecorrection}), ignoring the second scaling term $f_2$ (or equivalently using Eq.~(\ref{escale1}) taking $w=0$). To carry out such a fitting we want to use data that is ``close enough" to the critical point, i.e. small enough $\delta\phi$ and $\dot\gamma$, so as to be the scaling region.  However we also need a parametrization of the unknown scaling function $f_1(x)$.  Because of the wide range of values spanned by $p$ and $\sigma$, we choose an exponential parametrization, using,
\begin{equation}
f_1(x)=\mathrm{exp}\left({\sum_{n=0}^5 a_n x^n}\right),
\label{escalfun}
\end{equation}
and thus fit our data to $p,\sigma=\dot\gamma^q f_1([\phi-\phi_J]/\dot\gamma^{1/z\nu})$, with $\phi_J$, $q$, $1/z\nu$, and $a_0$ to $a_5$ as free fitting parameters.  We use the Levenberg--Marquardt algorithm to do our fitting.

Because our parametrization in Eq.~(\ref{escalfun}) involves an expansion in $x$ to finite (i.e. fifth) order, it will be an acceptable form for fitting only at sufficiently {\em small} $x$.  We have therefore concentrated our efforts on simulations where $x$ is suitably small, as indicated in Fig.~\ref{params}.  However, it is important to realize that the scaling form of Eq.~(\ref{ecorrection}) is valid at all values of $x=\delta\phi/\dot\gamma^{1/z\nu}$, provided that $\phi$ and $\dot\gamma$ are both sufficient close to the critical point. Thus, once we have determined values for $\phi_J$, $q=y/z\nu$, and $1/z\nu$ from fits at small $x$, then plotting our data as in Eq.~(\ref{ecollapse}) should give a good collapse even for data points with larger values of $x$ outside the fitting region, provided the data points $(\phi,\dot\gamma)$ are all {\em sufficiently close} to the jamming critical point $(\phi_J, 0)$.

To determine the goodness of our fits, we measure the chi squared per degree of freedom, $\chi^2_\mathrm{dof}$.  Fits are judged to be reasonably good when $\chi^2_\mathrm{dof}\sim O(1)$.  
We carry out fits to $p$ and $\sigma$ separately, using only data with $0.838\le\phi\le 0.846$, within 0.6\% of $\phi_J$.  We have confirmed that restricting the data to a narrower window in $\phi$ does not change our results.  Since we do find that our results are quite sensitive to the range of $\dot\gamma$ used in the fit, we
systematically restrict the data to $\dot\gamma\le \dot\gamma_\mathrm{max}$, using decreasing values of $\dot\gamma_\mathrm{max}$, in order to control how close our data are to the critical point $\dot\gamma\to 0$.
We also restrict the data to values where $|x|=|\delta\phi|/\dot\gamma^{1/z\nu}\le x_\mathrm{max}$, in order to test over how wide a range of $x$ our parametrization of the scaling function in Eq.~(\ref{escalfun}) will be reasonable \cite{xmax}.  We then study how the results of our fits depend on the  cutoffs $\dot\gamma_\mathrm{max}$ and $x_\mathrm{max}$.

\begin{figure}[h!]
\includegraphics[width=3.2in]{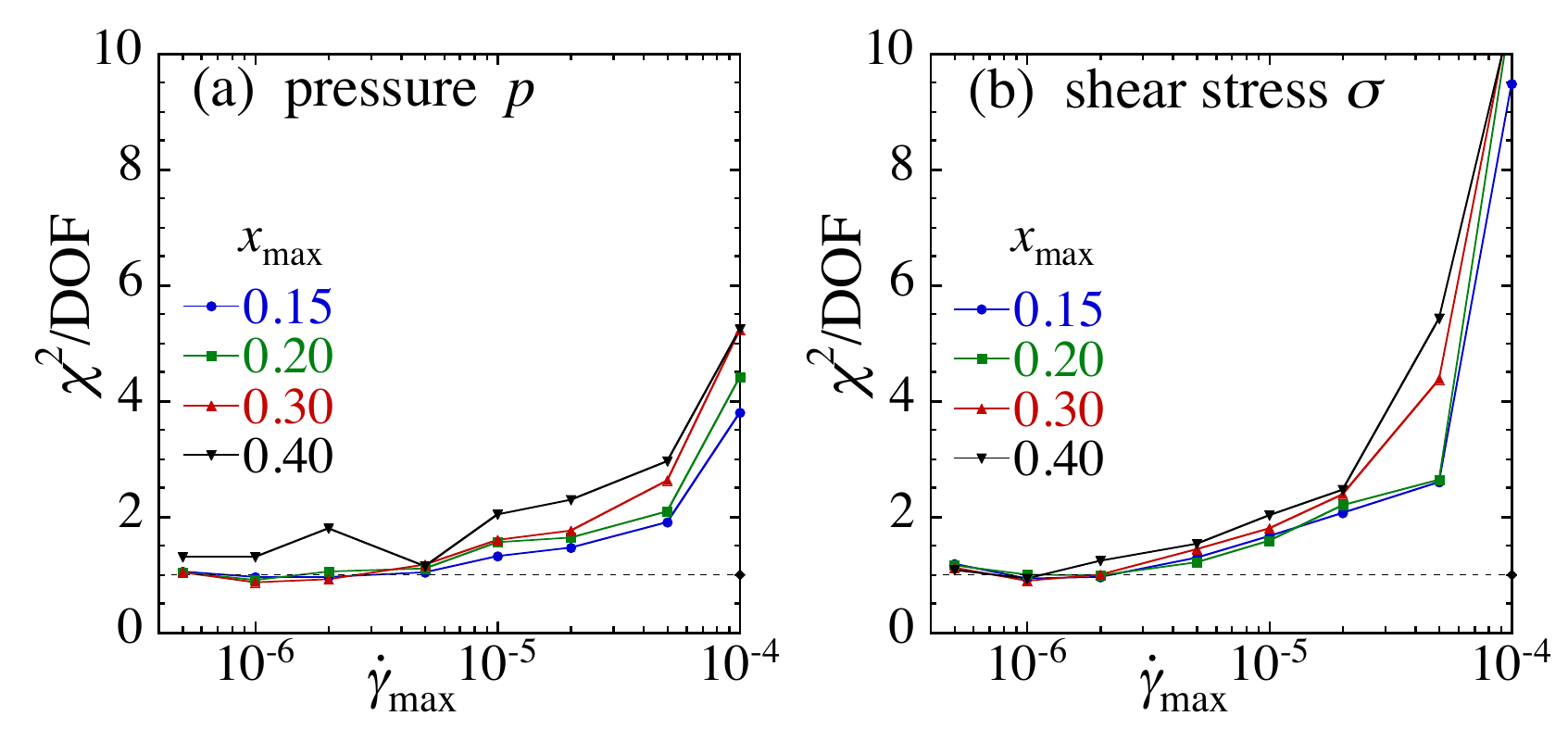}
\caption{(Color online) Chi squared per degree of freedom, $\chi^2_\mathrm{dof}$, of our fits of (a) pressure $p$, and (b) shear stress $\sigma$ to the scaling form of Eq.~(\ref{escale1}) without corrections-to-scaling (i.e. taking $w=0$), as a function of the upper limit $\dot\gamma_\mathrm{max}$ of data used in the fit.  We show results for several different values of $x_\mathrm{max}$, where only data with $|x|=|\delta\phi|/\dot\gamma^{1/z\nu}\le x_\mathrm{max}$ are used in the fit.   Data are restricted to the range $0.838\le\phi\le 0.846$.
}
\label{chisq-p-s}
\end{figure}

\begin{figure}[h!]
\includegraphics[width=3.2in]{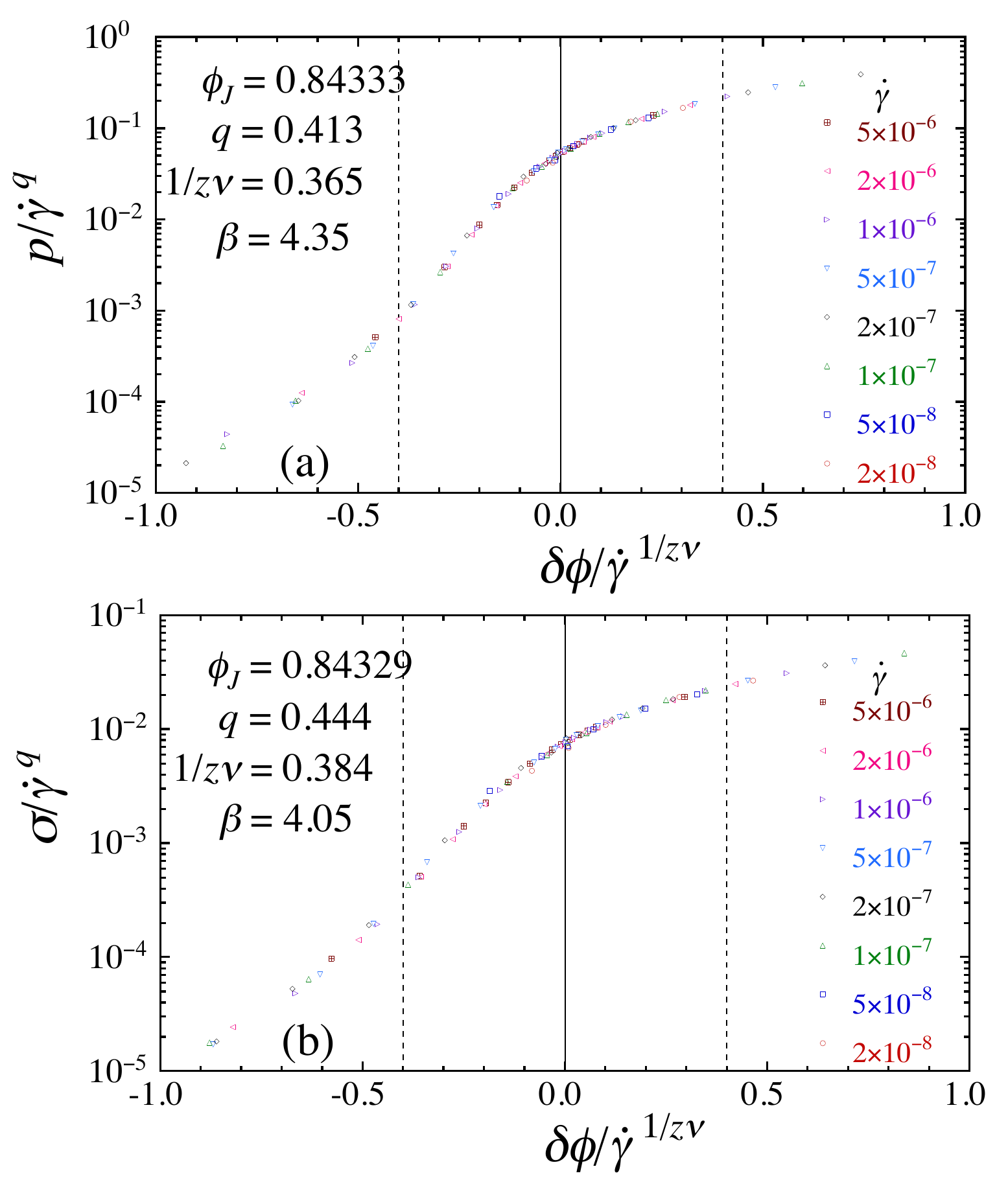}
\caption{(Color online)   Scaling collapse of (a) pressure, and (b) shear stress $\sigma$, plotted as $p/\dot\gamma^q$ and $\sigma/\dot\gamma^q$ vs $x=\delta\phi/\dot\gamma^{1/z\nu}$.  Data are restricted to the ranges $0.838\le\phi\le 0.846$ and $\dot\gamma\le5\times 10^{-6}$.  Only data for $|x|\le 0.4$, i.e. the data between the two vertical dashed lines, were used in doing the fit to the scaling function, however data at any value of $x$ are shown in the plot.  The resulting fitted values of $\phi_J$, $q$, and $1/z\nu$, as well as the exponent $\beta = (2-q)z\nu$, are as shown in the figures.
}
\label{p-s-scaled}
\end{figure}

In Fig.~\ref{chisq-p-s} we show the resulting $\chi^2_\mathrm{dof}$ for our fits to $p$ and to $\sigma$, vs the strain rate cutoff $\dot\gamma_\mathrm{max}$, for several different values of $x_\mathrm{max}$. We see that the fits look reasonable, i.e. $\chi^2_\mathrm{dof}\sim 1$, when $\dot\gamma\le 5\times 10^{-6}$, and $|x|\le 0.4$.  We therefore use the values of $\phi_J$, $q$, and $1/z\nu$ obtained from the fits using $\dot\gamma_\mathrm{max}=5\times10^{-6}$ and $x_\mathrm{max}=0.4$, and in Fig.~\ref{p-s-scaled} show the resulting data collapses for $p$ and for $\sigma$, according to Eq.~(\ref{ecollapse}).  Only our data satisfying $0.838\le\phi\le 0.846$ and $\dot\gamma\le 5\times 10^{-6}$ are plotted.
Even though only data with $|x|=|\delta\phi|/\dot\gamma^{1/z\nu}\le 0.4$ were used in generating the fit, all the data for $|x|\le 1$ appear to collapse reasonably well to the same continuous curve.  The fitted values of $\phi_J$, $q$, and $1/z\nu$ that were used to obtain these collapses, as well as the exponent $\beta = (2-q)z\nu$, are indicated in the figures.

Although the data collapses in Fig.~\ref{p-s-scaled} appear quite good to the eyeball, and although the fits are quantitatively good with $\chi^2_\mathrm{dof}\sim 1$, it is somewhat troubling that the fitted exponents for $p$ do not agree with those for $\sigma$, as we would have expected (and as we found earlier in a model with Newtonian rheology \cite{OlssonTeitelScaling}).  In particular, from $p$ we find $\beta=4.35\pm 0.04$, while from $\sigma$ we find $\beta=4.05\pm 0.04$; thus the two values of $\beta$ are not equal within the estimated statistical errors.  That there is a problem becomes more apparent if we look at the dependence of the fitted parameters on the values of the fit cutoffs $\dot\gamma_\mathrm{max}$ and $x_\mathrm{max}$.  To have a stable self-consistent fit, we need not only $\chi^2_\mathrm{dof}\sim 1$, but also that {\em the fitted parameters remain constant, within the estimated statistical errors, as the window of fitted data shrinks closer to the critical point}, i.e. as $\dot\gamma_\mathrm{max}$ decreases.

\begin{figure}[h!]
\includegraphics[width=3.2in]{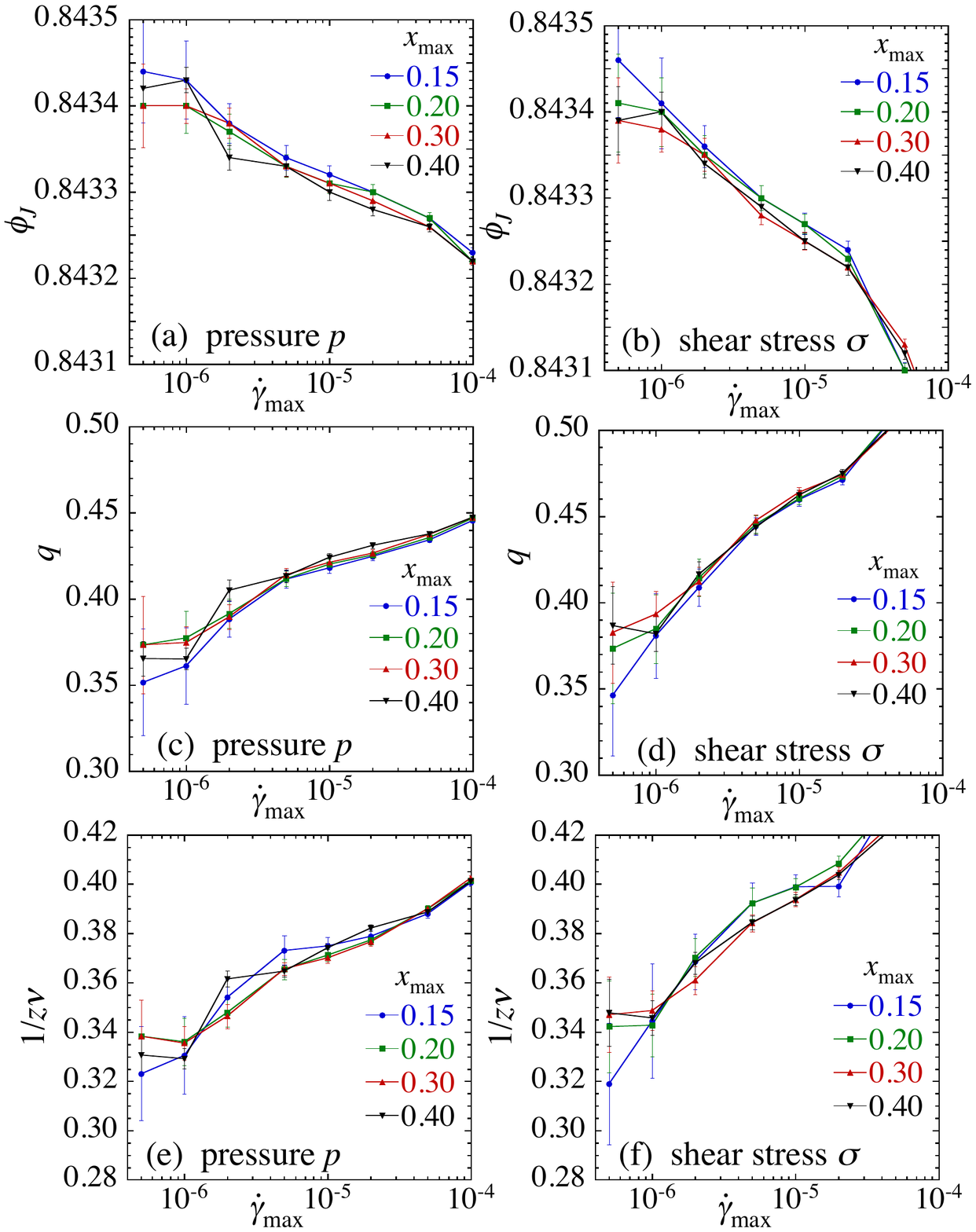}
\caption{(Color online)  Fitted parameters $\phi_J$, $q$, and $1/z\nu$, for  pressure $p$ (left column), and shear stress $\sigma$ (right column), vs the strain rate cutoff $\dot\gamma_\mathrm{max}$ that defines the range of data, $\dot\gamma\le\dot\gamma_\mathrm{max}$, used in the fit.  We show results for different values of the additional cutoff $x_\mathrm{max}$, where only data with $|x|=|\delta\phi|/\dot\gamma^{1/z\nu}\le x_\mathrm{max}$ are used in the fit.  Results are from fits to the scaling form of Eq.~(\ref{escale1}) {\em without} corrections-to-scaing (i.e. taking $w=0$).
}
\label{fitparams}
\end{figure}

In Fig.~\ref{fitparams} we show the fit parameters  $\phi_J$, $q$ and $1/z\nu$ that result when we fit $p$ and $\sigma$ separately to the scaling form of Eq.~(\ref{escale1}) (with $w=0$), restricting the data used in the fit to $\dot\gamma\le\dot\gamma_\mathrm{max}$ and $|x|=|\delta\phi|/\dot\gamma^{1/z\nu}\le x_\mathrm{max}$.  We plot the parameters vs $\dot\gamma_\mathrm{max}$ for several different values of $x_\mathrm{max}$.  We see that there is little significant dependence on the choice of $x_\mathrm{max}$, however there is a clear and systematic dependence on the value of $\dot\gamma_\mathrm{max}$.  In particular, $\phi_J$ systematically increases, and $q$ and $1/z\nu$ systematically decrease, as $\dot\gamma_\mathrm{max}$ decreases; this remains true even for $\dot\gamma\le 5\times 10^{-6}$ where the $\chi^2_\mathrm{dof}$ has become roughly equal to unity.  

\begin{figure}[h!]
\includegraphics[width=3.2in]{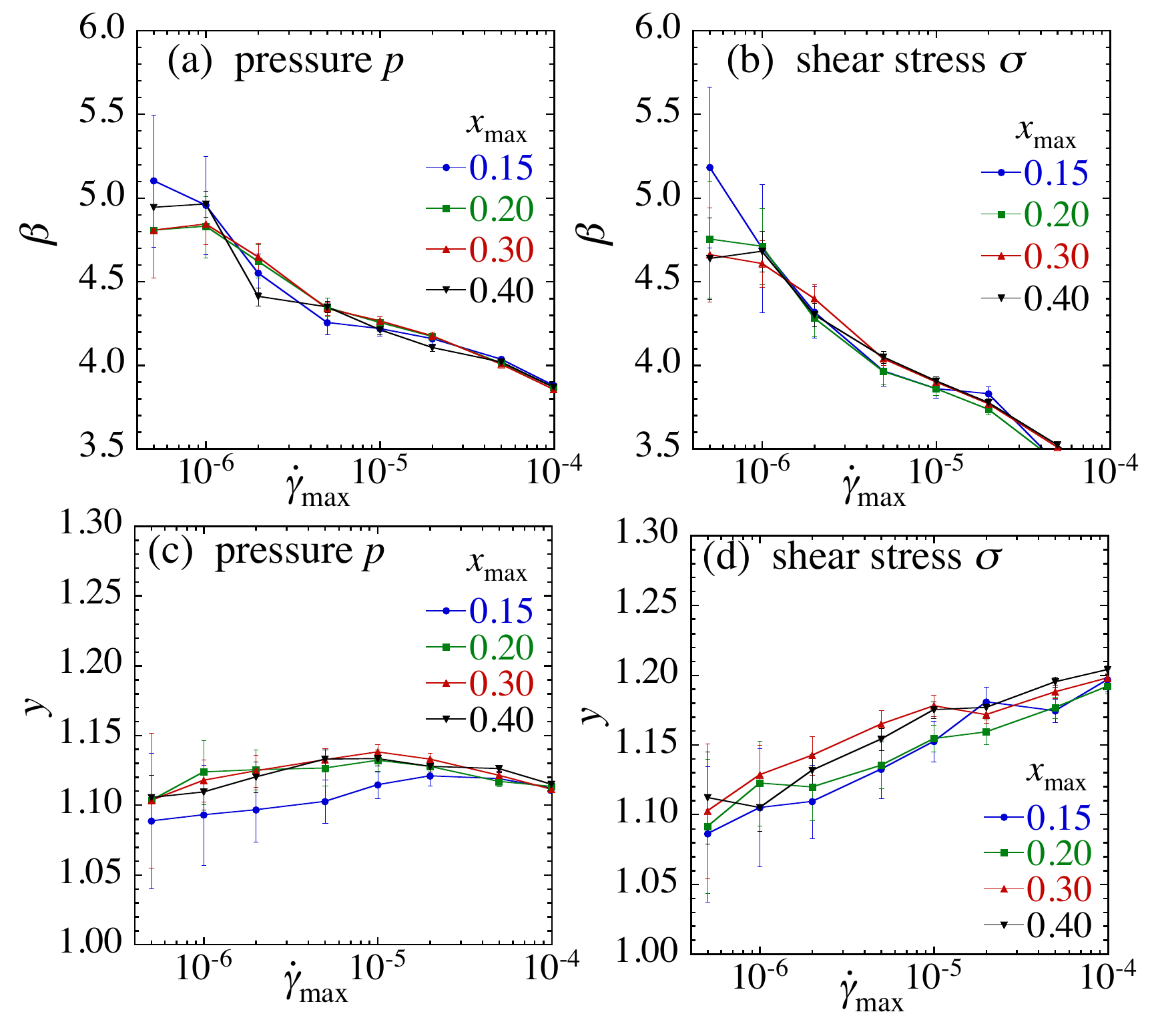}
\caption{(Color online)  Exponents $\beta=(2-q)z\nu$ and $y=qz\nu$, obtained from the fit parameters of Fig.~\ref{fitparams} for  pressure $p$ (left column), and shear stress $\sigma$ (right column), vs the strain rate cutoff $\dot\gamma_\mathrm{max}$.  We show results for different values of the additional cutoff $x_\mathrm{max}$, where only data with $|x|=|\delta\phi|/\dot\gamma^{1/z\nu}\le x_\mathrm{max}$ are used in the fit.  Results are from fits to the scaling form of Eq.~(\ref{escale1}) {\em without} corrections-to-scaing (i.e. taking $w=0$).
}
\label{fitparams2}
\end{figure}

In Fig.~\ref{fitparams2} we similarly show the exponents for the Bagnold transport coefficient and the yield stress, $\beta=(2-q)z\nu$ and $y=qz\nu$, vs $\dot\gamma_\mathrm{max}$ for different $x_\mathrm{max}$.
The behavior of $\beta$ that we see here is consistent with the behavior observed in previous simulations, as discussed in Sec.~\ref{sPrevVol}, in that $\beta$ increases as $\dot\gamma_\mathrm{max}$ decreases, and we find similar numerical values for $\beta$ when considering the larger values of $\dot\gamma_\mathrm{max}$ that were used in these earlier works.  For all the parameters $\phi_J$, $q$, $1/z\nu$, $\beta$ and $y$, we see that in general the values obtained from the fits to $p$ appear to be agreeing with those obtained from the fits to $\sigma$ only at the smallest values of $\dot\gamma_\mathrm{max}$; at the larger $\dot\gamma_\mathrm{max}$ they can be quite noticeably different.

Thus, while the fits are quantitatively good, and the scaling collapses of Fig.~\ref{p-s-scaled} appear to be good,  they do not give self-consistent results in that the values of the fit parameters are continuously changing as $\dot\gamma_\mathrm{max}$ decreases.
This leads us to conclude that our simple approach in this section, ignoring the leading correction-to-scaling, is not adequate for describing the critical behavior of the rheology over the range of parameters we have simulated.

\subsection{Including corrections-to-scaling}
\label{ssCorrections}

Since the approach of the previous section failed to give consistent results, we now reanalyze our data by including the leading correction-to-scaling according to Eq.~(\ref{ecorrection}).  Because of the $\dot\gamma^{\omega/z}$ prefactor of the scaling function $f_2(x)$ in Eq.~(\ref{ecorrection}), when the correction-to-scaling term is no longer negligible there can be no nice scaling collapse of the data when plotted according to Eq.~(\ref{ecollapse}).  

We may still, however, get a graphical sense of the effect of the correction-to-scaling by considering the following.  In the limit of $\dot\gamma\to 0$ we expect the following behaviors for the pressure $p$ (and similarly for the shear stress $\sigma$): (i) below $\phi_J$, $p$ vanishes as $p\propto\dot\gamma^2$, (ii) above $\phi_J$, $p\to p_0$ the finite yield stress, and (iii) exactly at $\phi_J$, $p\propto \dot\gamma^q$.  If we now consider the quantity $p/\dot\gamma^q$, we expect that (i) below $\phi_J$, $p/\dot\gamma^q$ vanishes as $p/\dot\gamma^q\propto\dot\gamma^{2-q}$, (ii) above $\phi_J$, $p/\dot\gamma^q$ diverges as $p/\dot\gamma^q\propto \dot\gamma^{-q}$, and (iii) exactly at $\phi_J$, $p/\dot\gamma^q$ is constant.  If we now consider the behavior at $\phi_J$ as $\dot\gamma$ increases, then $p/\dot\gamma^q$ will depart from the  limiting small $\dot\gamma$ constant when the correction-to-scaling term $\sim \dot\gamma^{\omega/z}$ becomes non-negligible.  

\begin{figure}[h!]
\includegraphics[width=3.2in]{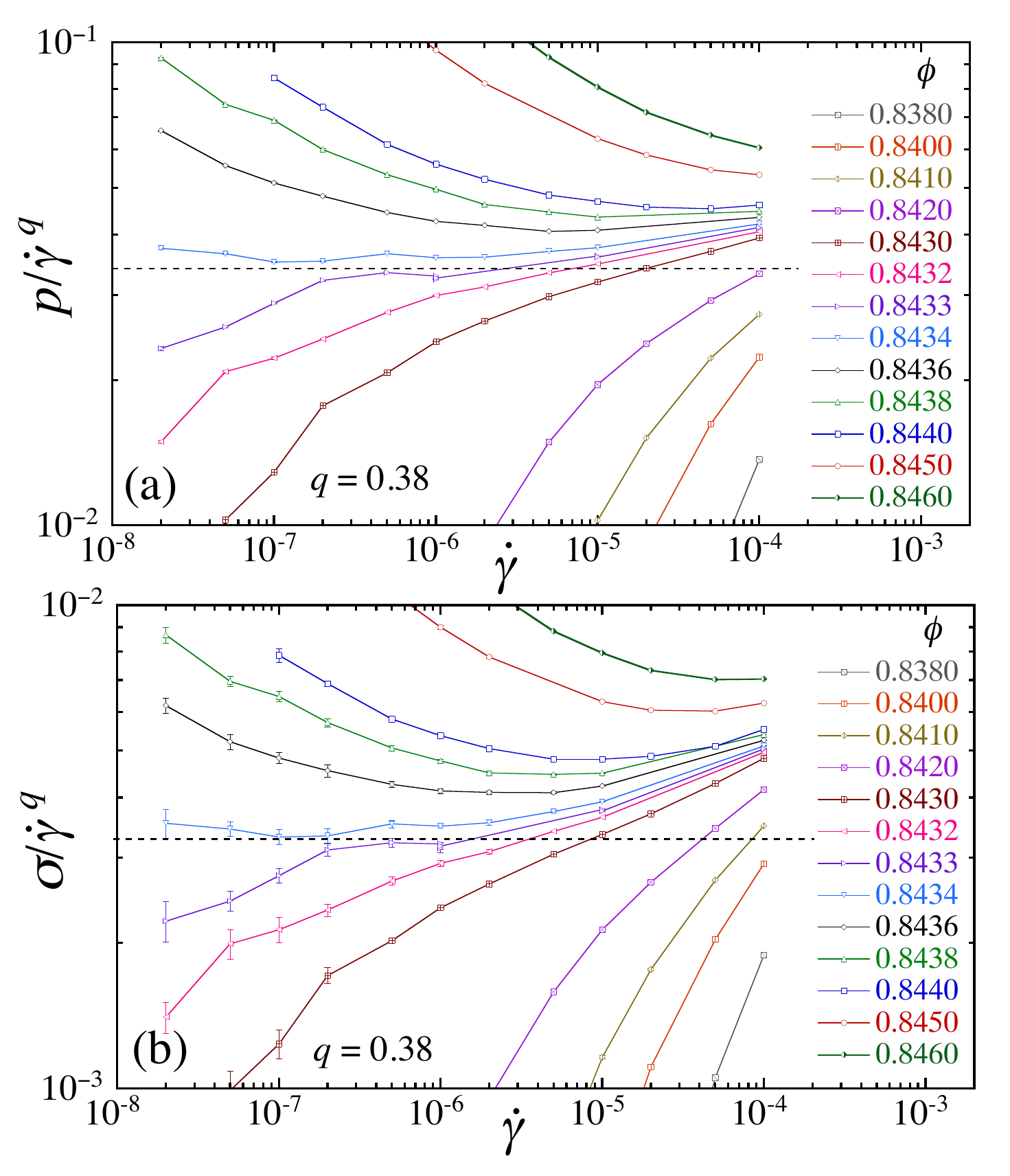} 
\caption{(Color online)  (a) $p/\dot\gamma^q$ and (b) $\sigma/\dot\gamma^q$ vs the strain rate $\dot\gamma$ for different values of the packing fraction $\phi$.  The value $q=0.38$, obtained from our scaling analysis, is used.  The dashed lines represent the small $\dot\gamma$ limiting values exactly at $\phi_J$, and separate the curves with $\phi>\phi_J$ (above the dashed line at small $\dot\gamma$) from those with $\phi<\phi_J$ (below the dashed line at small $\dot\gamma$).  The value of $\phi$ decreases as curves go from top to bottom.
}
\label{p-sogdotq}
\end{figure}

In Fig.~\ref{p-sogdotq} we plot $p/\dot\gamma^q$ and $\sigma/\dot\gamma^q$, using the value $q=0.38$ as found by our subsequent analysis detailed below.  We see that $\phi_J\approx 0.84335$ separates the curves that curve upwards as $\dot\gamma$ decreases (these are above $\phi_J$) from the curves that curve downwards (these are below $\phi_J$).  The dashed lines represent the constant values of $p/\dot\gamma^q$ and $\sigma/\dot\gamma^q$ expected at $\phi_J$ for sufficiently small $\dot\gamma$.  If we look at the curves closest to $\phi_J$, i.e. at $\phi=0.8433$ and $0.8434$, we see that they are roughly flat for a wide range of $\dot\gamma$, and then curve upwards as $\dot\gamma$ increases; this is the effect of the $\dot\gamma^{\omega/ z}$ correction-to-scaling term.  Comparing $p$ to $\sigma$ in Fig.~\ref{p-sogdotq}, we see that the correction-to-scaling term is larger for the shear stress $\sigma$ than for the pressure $p$; a similar conclusion was previously found for a related model system with Newtonian rheology \cite{OlssonTeitelScaling}.  Our results of Fig.~\ref{p-sogdotq} emphasize that the rheology $p,\sigma\sim\dot\gamma^q$ expected exactly at $\phi_J$, only holds asymptotically at sufficiently small $\dot\gamma$, and does not persist to arbitrarily large values of $\dot\gamma$.

We now fit our data to the scaling form of Eq.~(\ref{ecorrection}) and test whether the fit is good, i.e. $\chi^2_\mathrm{dof}\sim 1$, and whether the values of the fitted parameters remain consistent as we vary the window of data used in the fit.
To carry out this data fitting we parametrize the two scaling functions of Eq.~(\ref{ecorrection}) as,
\begin{equation}
f_1(x)=\mathrm{exp}\left(\sum_{n=0}^4a_nx^n\right),\>
f_2(x)=b_0\mathrm{exp}\left(\sum_{n=1}^3b_nx^n\right).
\label{escalfun2}
\end{equation}
In contrast to the previous section, here we use an expansion of lower order in $x$ in order to keep the total number of fit parameters manageable.  We thus might expect, and indeed we do find, that our fitting will be more sensitive to the choice of $x_\mathrm{max}$ than was found in the previous section. 
Since the correction-to-scaling term needs to be sizable if we are to determine it properly, here
we choose our smallest $\dot\gamma_\mathrm{max}=5\times 10^{-6}$, larger than the value $5\times 10^{-7}$ used  in the previous section.
For $f_2$ we include the multiplicative factor $b_0$, rather than writing it as $\mathrm{exp}(b_0)$ as in $f_1$, since we wish to allow for the possibility that the correction term could be negative; in practice, however, we always find that $b_0>0$.

\begin{figure}[h!]
\includegraphics[width=3.2in]{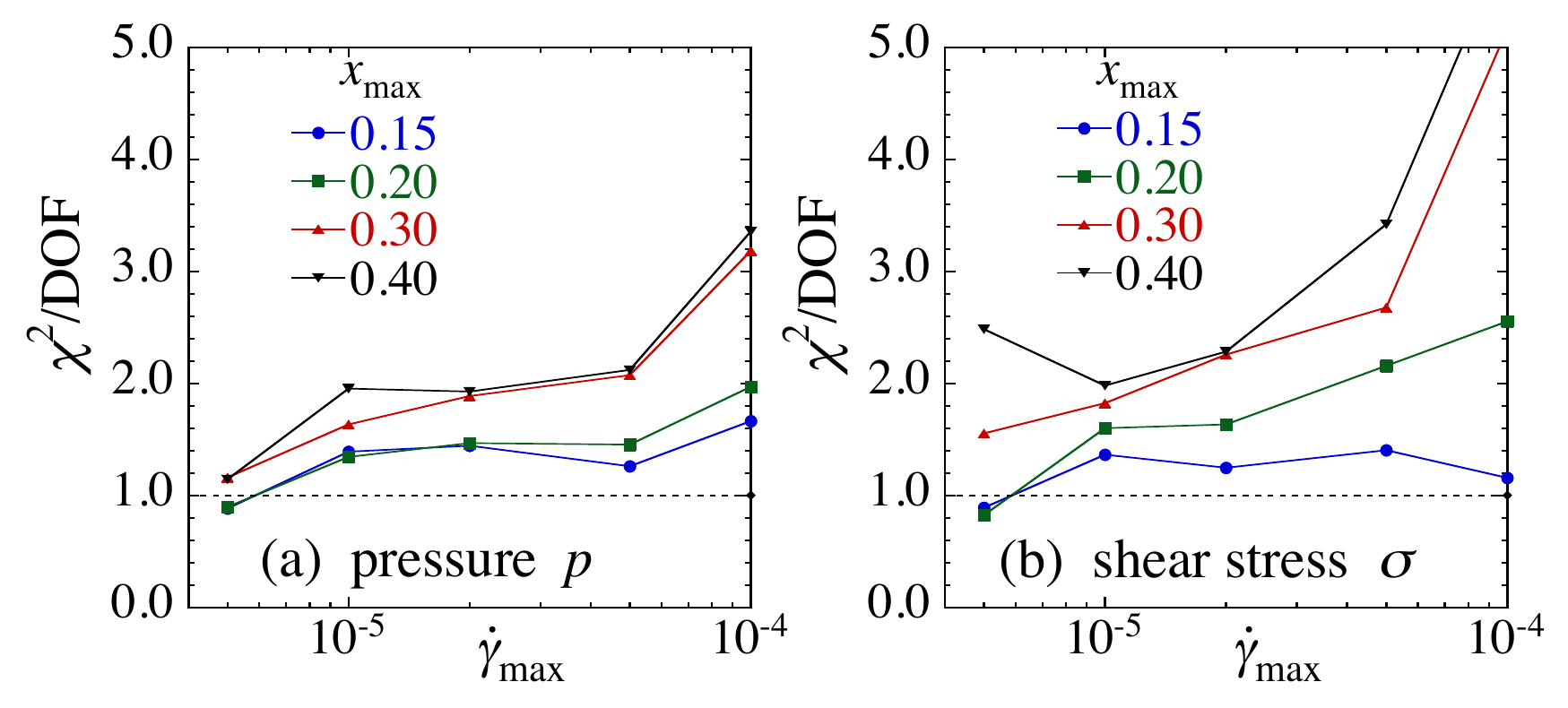}
\caption{(Color online) Chi squared per degree of freedom, $\chi^2_\mathrm{dof}$, of our fits of (a) pressure $p$, and (b) shear stress $\sigma$ to the scaling form of Eq.~(\ref{ecorrection}) including the correction-to-scaling, as a function of the upper limit $\dot\gamma_\mathrm{max}$ of data used in the fit.  We show results for several different values of $x_\mathrm{max}$, where only data with $|x|=|\delta\phi|/\dot\gamma^{1/z\nu}\le x_\mathrm{max}$ are used in the fit.  
}
\label{chi-p-s-corr}
\end{figure}

In Fig.~\ref{chi-p-s-corr} we show the $\chi^2_\mathrm{dof}$ for such fits to $p$ and $\sigma$ separately, as a function of $\dot\gamma_\mathrm{max}$ for several different values of $x_\mathrm{max}$.  The fits seem reasonable, with $\chi^2_\mathrm{dof}\lesssim1.5$, for all $\dot\gamma_\mathrm{max}\le 5\times 10^{-5}$ at the two smallest $x_\mathrm{max}$.

\begin{figure}[h!]
\includegraphics[width=3.2in]{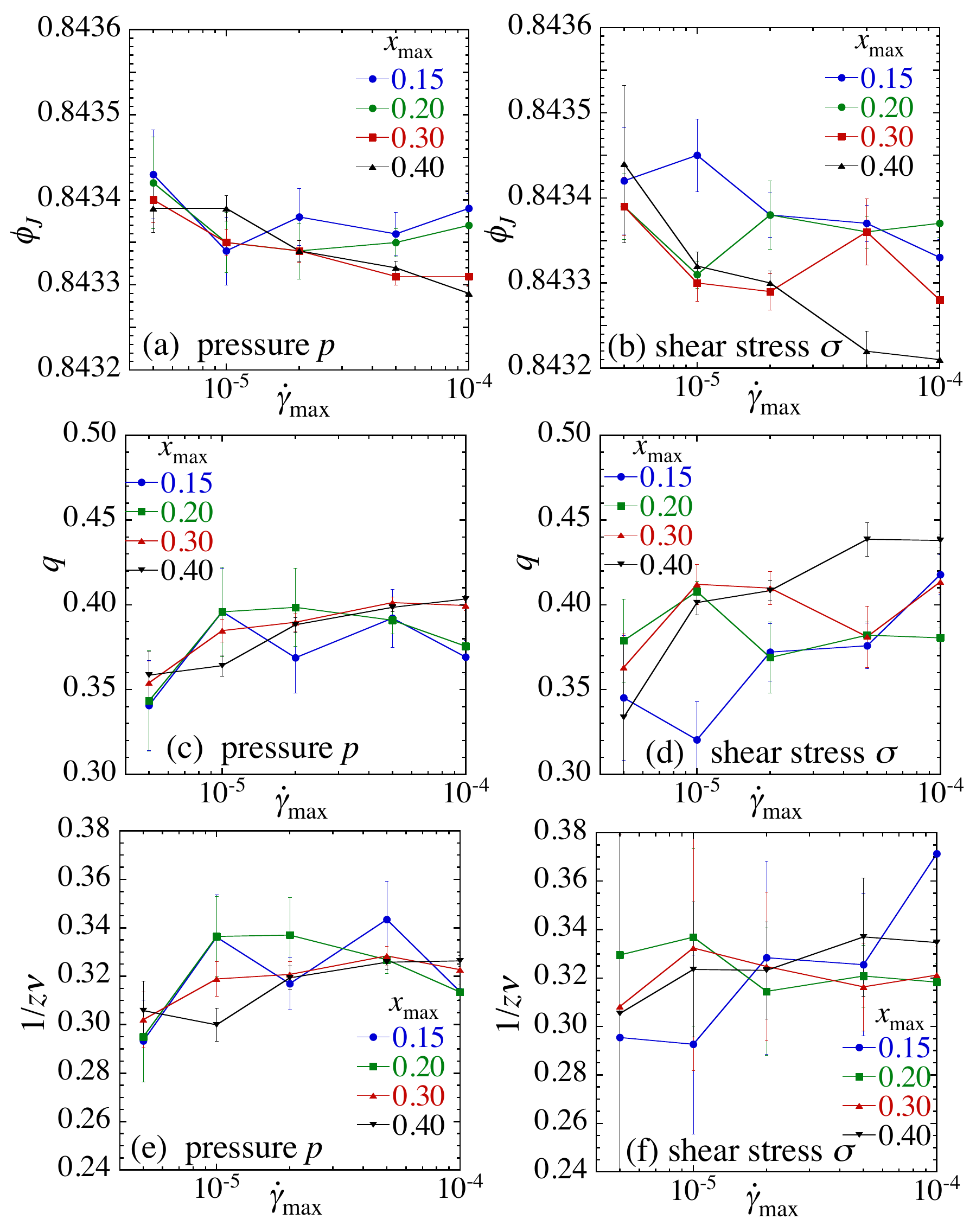}
\caption{(Color online)  Fitted parameters $\phi_J$, $q$, and $1/z\nu$, for  pressure $p$ (left column), and shear stress $\sigma$ (right column), vs the strain rate cutoff $\dot\gamma_\mathrm{max}$ that defines the range of data, $\dot\gamma\le\dot\gamma_\mathrm{max}$, used in the fit.  We show results for different values of the additional cutoff $x_\mathrm{max}$, where only data with $|x|=|\delta\phi|/\dot\gamma^{1/z\nu}\le x_\mathrm{max}$ are used in the fit.  Results are from fits to the scaling form of Eq.~(\ref{ecorrection}) {\em including} corrections-to-scaling.
}
\label{fitparams-corr}
\end{figure}

\begin{figure}[h!]
\includegraphics[width=3.2in]{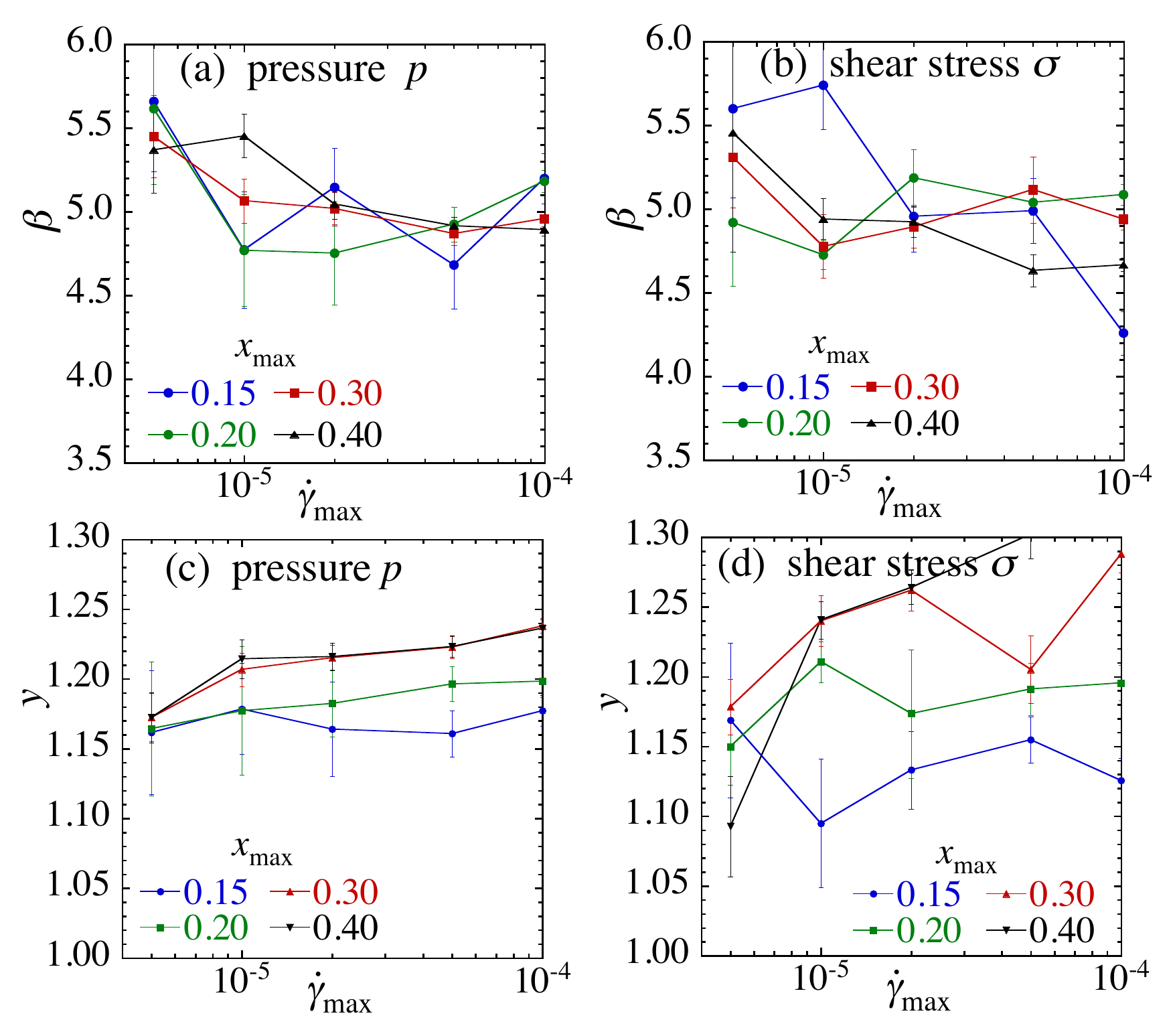}
\caption{(Color online)  
Exponents $\beta=(2-q)z\nu$ and $y=qz\nu$, obtained from the fit parameters of Fig.~\ref{fitparams-corr} for  pressure $p$ (left column), and shear stress $\sigma$ (right column), vs the strain rate cutoff $\dot\gamma_\mathrm{max}$.  We show results for different values of the additional cutoff $x_\mathrm{max}$, where only data with $|x|=|\delta\phi|/\dot\gamma^{1/z\nu}\le x_\mathrm{max}$ are used in the fit.  Results are from fits to the scaling form of Eq.~(\ref{ecorrection}) {\em including} corrections-to-scaling.
}
\label{fitparams2-corr}
\end{figure}

In Fig.~\ref{fitparams-corr} we show the resulting fit parameters  $\phi_J$, $q$ and $1/z\nu$; we will consider the correction-to-scaling exponent $\omega/z$ in the following section.  
In Fig.~\ref{fitparams2-corr} we show the exponents $\beta=(2-q)z\nu$ and $y=qz\nu$, as computed from the exponent values shown in Fig.~\ref{fitparams-corr}.
We plot these parameters vs $\dot\gamma_\mathrm{max}$ for several different values of $x_\mathrm{max}$.  Compared to the corresponding results of Figs.~\ref{fitparams} and \ref{fitparams2} without corrections-to-scaling, here we see (i) no strong systematic dependence of the parameters on $\dot\gamma_\mathrm{max}$,
(ii) greater consistency comparing $p$ and $\sigma$ over the entire range of $\dot\gamma_\mathrm{max}$ (in Fig.~\ref{fitparams} parameters tend to agree only at the smaller $\dot\gamma_\mathrm{max}$, but not at the larger $\dot\gamma_\mathrm{max}$), (iii) greater sensitivity to the choice of $x_\mathrm{max}$, particularly for $\sigma$, and (iv) larger statistical errors which may be attributed to the increase in the number of fitting parameters, and to the loss of accuracy in the fitting functions at larger values of $x$ (because the expansion in powers of $x$ is truncated at lower order; compare Eqs.~(\ref{escalfun}) and (\ref{escalfun2})).

We thus find that the fit of our data to the scaling form of Eq.~(\ref{ecorrection}), including the corrections-to-scaling, gives a reasonable fit with  consistent values for the fitting parameters; however the accuracy of these parameters suffers from the effects described in (iv) above.  We conclude that $\phi_J\approx 0.84335\pm0.00010$, $q\approx 0.38\pm0.05$, $1/z\nu\approx 0.32\pm 0.02$, $\beta\approx 5.0\pm 0.4$ and $y\approx 1.15\pm 0.05$.
We note that these values are consistent (within the estimated errors) with the results found from our fits ignoring the correction-to-scaling, shown in Figs.~\ref{fitparams} and \ref{fitparams2}, provided we consider in those figures only the smallest value of $\dot\gamma_\mathrm{max}$.  This thus suggests that the correction-to-scaling term is becoming negligible at the smallest strain rates $\dot\gamma$ that we simulate.  In Table I we compare the values of the exponents found in the present work for Bagnoldian rheology, with the corresponding exponents found in Ref. \cite{OlssonTeitelScaling} for Newtonian rheology.  We see that the values of $\phi_J$ and the exponent $y$ agree within the estimated errors, and that the value of $y$ is slightly bigger than unity; however the exponents $q$ and $1/z\nu$ appear to be different for the two different rheologies.

We note that the value of $q$ found here for Bagnold rheology is in rough agreement with the value $q=2/5$ obtained from Otsuki and Hayakawa's phenomenological mean-field theory \cite{OH1,OH2}. However our value of $1/z\nu\approx 0.32$ is noticeably different from their value of $2/5$.  Thus our result for $\beta=2z\nu-y\approx 5$ is
clearly larger than the value of 4 predicted by Otsuki and Hayakawa \cite{OH1,OH2}, but is in agreement (within the estimated errors) with the numerical result of Peyneau and Roux \cite{Roux}.  Comparing our $\beta_\mathrm{Bagnold}$ with our previously determined $\beta_\mathrm{Newton}$ \cite{OlssonTeitelScaling}, we find that the prediction of DeGiuli et al. \cite{DeGiuli} that $\beta_\mathrm{Bagnold}=2\beta_\mathrm{Newton}$ is obeyed within the outer range of our error estimates, however our $\beta_\mathrm{Bagnold}=5.0\pm0.4$ is somewhat smaller than the value $5.7$ that one gets from their  calculation of the dilatancy exponent $a\approx 0.35$.  


\begin{table*}[htdp]
\caption{Critical parameters for Bagnoldian rheology as found in the present work, compared to the corresponding parameters for Newtonian rheology as found in Ref. \cite{OlssonTeitelScaling}, for frictionless particles with a harmonic elastic repulsion.  Here $n=2$ for Bagnoldian and $n=1$ for Newtonian rheology.}
\begin{center}
\begin{tabular}{|l|c|c|c|c|c|c|c|}
\hline
Model & $\phi_J$ & $q$ & $1/z\nu$ & $\beta=(n-q)z\nu$ & $y=qz\nu$ & $\omega/z$ &  $\omega\nu$ \\
\hline
Bagnoldian  & $0.84335\pm0.00010$ & $0.38\pm0.05$ & $0.32\pm 0.02$ & $5.0\pm 0.4$ & $1.15\pm 0.05$ & $0.35\pm0.07$  & $1.1\pm 0.3$   \\
Newtonian \cite{OlssonTeitelScaling} & $0.8435\pm 0.0002$ & $0.28\pm0.02$ & $0.26\pm 0.02$ & $2.8\pm 0.3$ & $1.08\pm 0.03$ &  $0.29\pm0.03$ & $1.10\pm 0.06$ \\
\hline
\end{tabular}
\end{center}
\label{default}
\end{table*}%

\subsection{Macroscopic friction}

In this section we discuss our results for the correction-to-scaling exponent, which is closely related to the
macroscopic friction, $\mu\equiv \sigma/p$.  From Eq.~(\ref{ea}) we have that the exponent $a$ of the dilatancy law Eq.~(\ref{ec1}) is $a=2/\beta\approx 0.4$.  From Eq.~(\ref{eb}) we then have that the exponent $b$ of the friction law Eq.~(\ref{ec2}) is related to the exponent $a$ by, $b/a=\omega\nu$, where $\omega$ is the correction-to-scaling exponent of Eq.~(\ref{ecorrection}).  The exponent combination $\omega\nu$ also gives the variation of the macroscopic friction $\mu$ with packing fraction $\phi$, as given in Eq.~(\ref{emuphi}).  We thus wish to determine $\omega\nu$.

\begin{figure}[h!]
\includegraphics[width=3.2in]{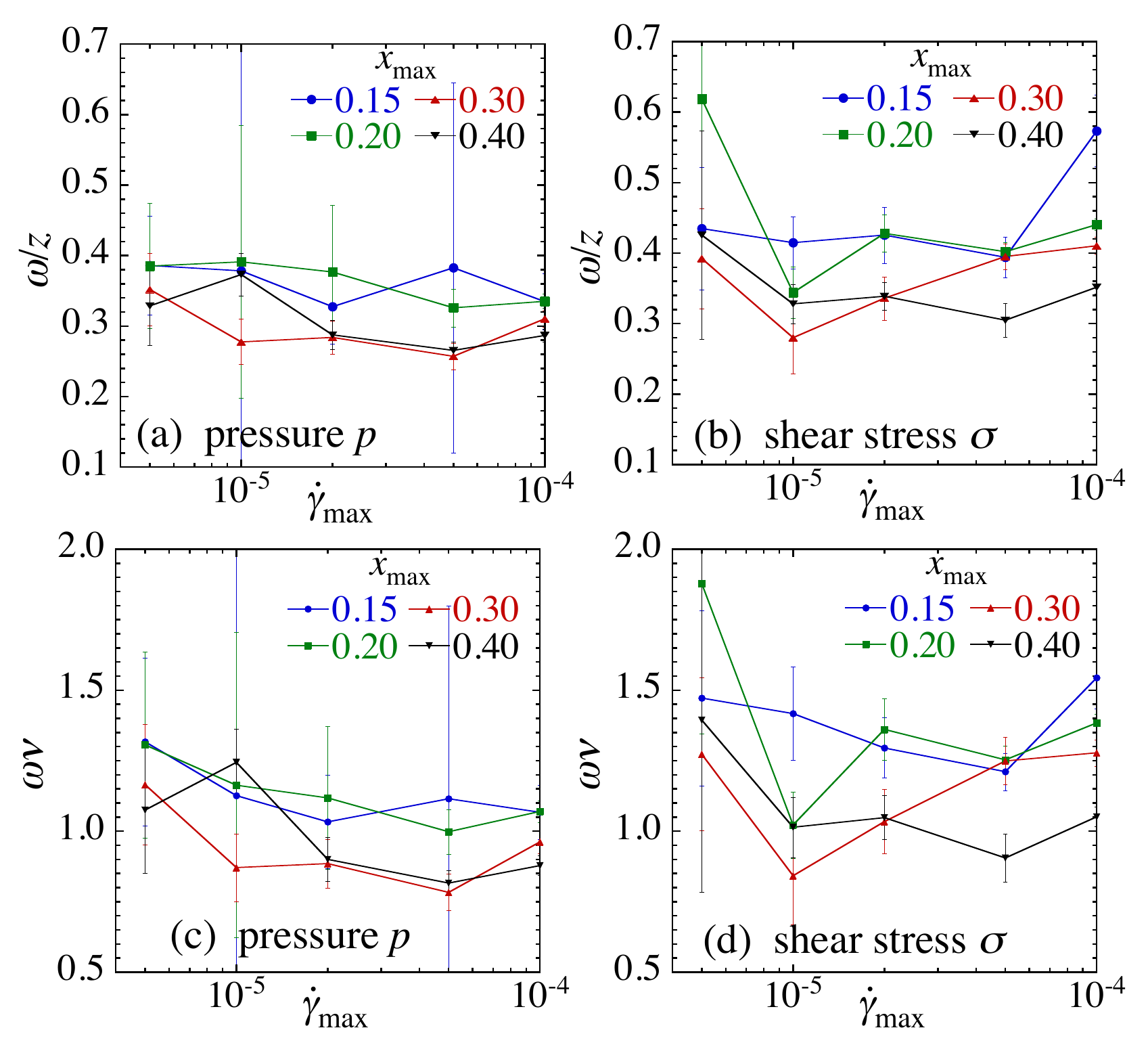}
\caption{(Color online) Correction-to-scaling exponent $\omega/z$, and the related exponent $\omega\nu$, as obtained from fits to pressure $p$ (left column),  and shear stress $\sigma$ (right column), vs the strain rate cutoff $\dot\gamma_\mathrm{max}$.  We show results for different values of the scaling parameter cutoff $x_\mathrm{max}$.  Only data with $\dot\gamma\le\dot\gamma_\mathrm{max}$ and  $|x|=|\delta\phi|/\dot\gamma^{1/z\nu}\le x_\mathrm{max}$ are used in the fit.  
}
\label{omeganu}
\end{figure}

Our fits to Eq.~(\ref{ecorrection}) determine the exponents $\omega/z$ and $1/z\nu$, from which we can then compute $\omega\nu=(\omega/z)/(1/z\nu)$.  In Fig.~\ref{omeganu} we plot the values of $\omega/z$ obtained from our fits to $p$ and to $\sigma$, and the resulting values of
$\omega\nu$, vs the strain rate cutoff $\dot\gamma_\mathrm{max}$ for several different values of $x_\mathrm{max}$.
We see that it is difficult to get accurate values of $\omega\nu$.  However our results are not  inconsistent with the value $\omega\nu=1$ claimed by DeGiuli et al. \cite{DeGiuli}, which was also found in the numerical simulations of Peyneau and Roux \cite{Roux}.    It is also consistent with the values $\omega\nu\approx 1$ that we previously found \cite{OlssonTeitelScaling,OTFSS} in a model with Newtonian rheology.

\begin{figure}[h!]
\includegraphics[width=3.2in]{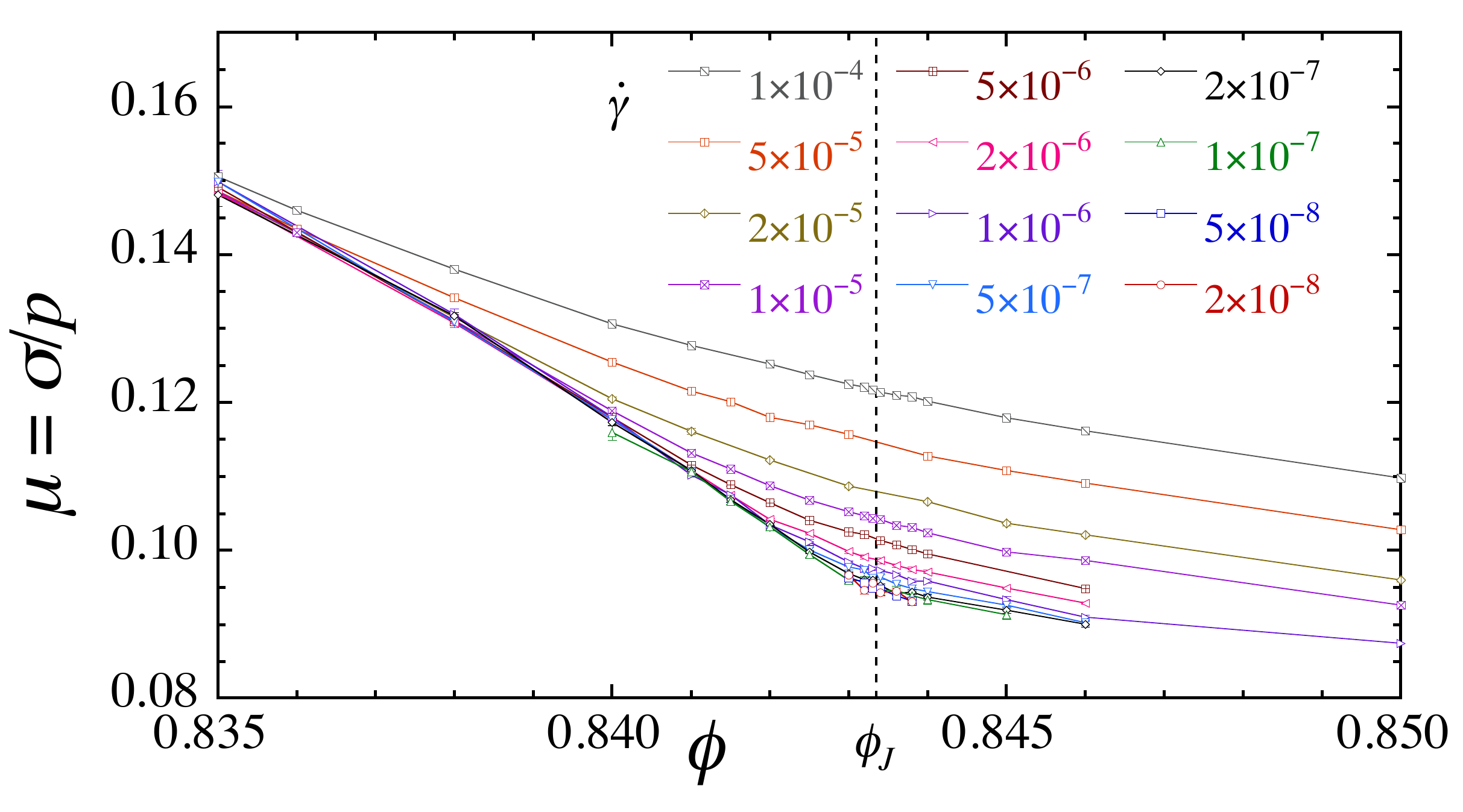}
\caption{(Color online) Macroscopic friction $\mu\equiv\sigma/p$ vs packing fraction $\phi$, for different values of the shear strain rate $\dot\gamma$.  The vertical dashed line locates the jamming transition at $\phi_J$.  The strain rate $\dot\gamma$ {\em decreases} as curves go from top to bottom. 
}
\label{sop}
\end{figure}

Our discussion of $\mu$ in Sec.~\ref{sConstituent} dealt specifically with the limit of hard-core particles below $\phi_J$.  We can, however, consider the more general case of $\mu$ for soft-core particles at finite $\dot\gamma$ and above $\phi_J$.  
In Fig.~\ref{sop} we show our results for $\mu$ vs $\phi$, for different values of $\dot\gamma$.  We see that as $\dot\gamma\to 0$, $\mu$ is everywhere approaching  a finite $\phi$-dependent constant. That $\mu$ is finite at $\phi_J$ as $\dot\gamma\to 0$ confirms, via Eq.~(\ref{ecorrection}),  that the scaling exponent $y$ (and hence $q$ and $\beta$) is the same for both $p$ and $\sigma$. 
A very similar looking plot of $\mu$ vs $\phi$ for models with Newtonian rheology was found in Ref.~\cite{VOT}.  

We can understand some of the features of our data in Fig.~\ref{sop} by considering the scaling form that $\mu$ should obey.  Since the exponents $q=y/z\nu$ are the same for $p$ and $\sigma$, we have from Eq.~(\ref{ecorrection}),
\begin{equation}
\mu\equiv\dfrac{\sigma}{p}= h_1\left(\dfrac{\delta\phi}{\dot\gamma^{1/z\nu}}\right)
+\dot\gamma^{\omega/z} h_2\left(\dfrac{\delta\phi}{\dot\gamma^{1/z\nu}}\right).
\end{equation}
Exactly at $\phi_J$, where $\delta\phi=0$, the above becomes,
\begin{equation}
\mu(\phi_J,\dot\gamma)=h_1(0)+h_2(0)\dot\gamma^{\omega/z}.
\label{esop-vs-gdot}
\end{equation}
Thus plotting $\mu$ at $\phi_J$ vs the strain rate $\dot\gamma$ should allow one to determine the correction-to-scaling exponent $\omega/z$.

\begin{figure}[h!]
\includegraphics[width=3.2in]{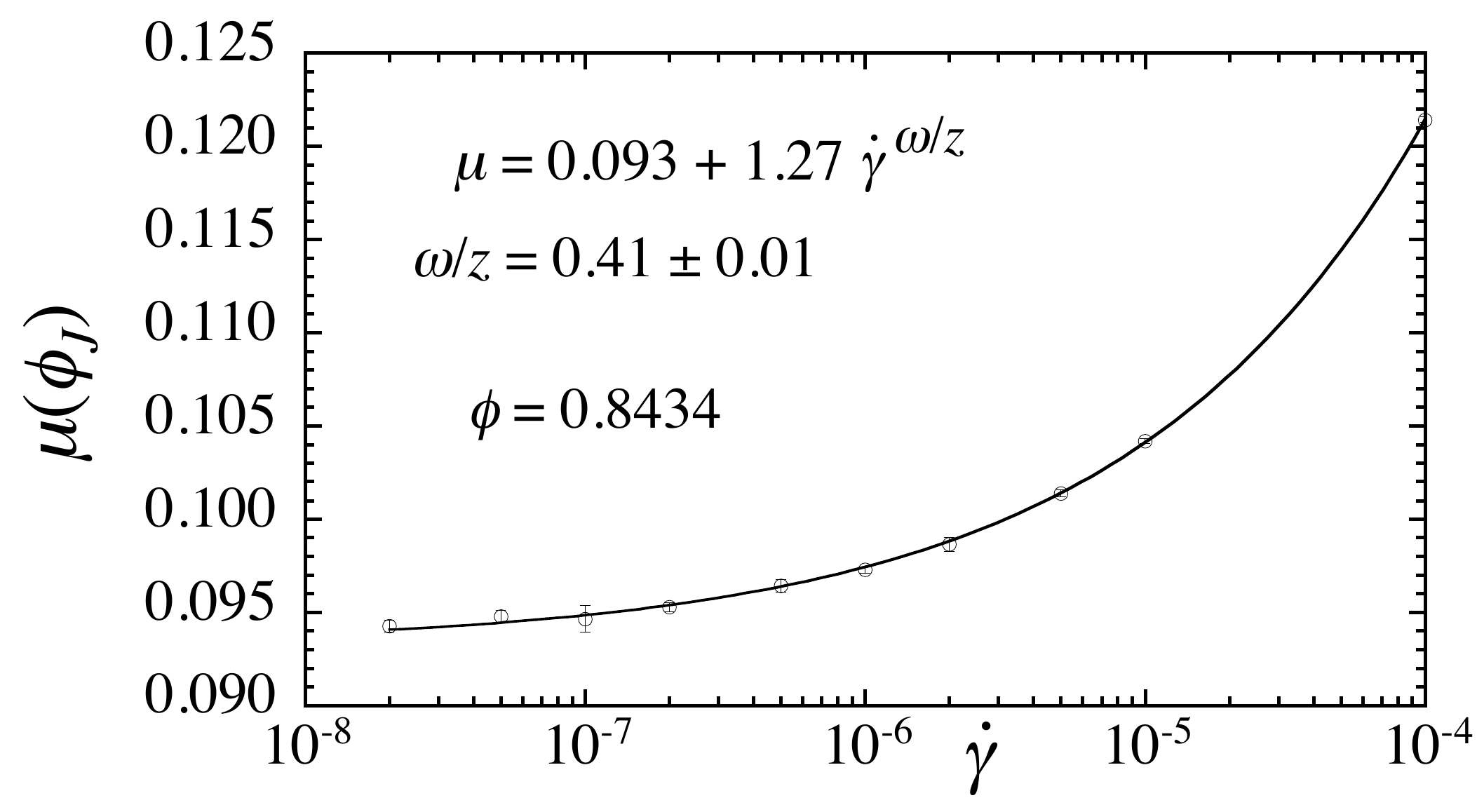}
\caption{Macroscopic friction $\mu\equiv\sigma/p$ vs shear strain rate $\dot\gamma$ at $\phi=0.8434\approx\phi_J$.  A fit to Eq.~(\ref{esop-vs-gdot}) determines the exponent $\omega/z\approx 0.41\pm 0.01$.
}
\label{sop-vs-gdot}
\end{figure}

In Fig.~\ref{sop-vs-gdot} we plot $\mu$ vs $\dot\gamma$ at $\phi=0.8434\approx\phi_J$.  Fitting to Eq.~(\ref{esop-vs-gdot}) we find the value $\omega/z\approx 0.41\pm 0.01$, consistent within the estimated errors with the results in Fig.~\ref{omeganu}.  We do not try any more elaborate fits to $\mu(\phi,\dot\gamma)$ since the quality of our data at the lowest $\dot\gamma$ is rather poor; the difference in values $\mu(\dot\gamma)-\mu(\dot\gamma^\prime)$, for neighboring values of $\dot\gamma$ and $\dot\gamma^\prime$, is less than the estimated errors on the values of $\mu(\dot\gamma)$ and $\mu(\dot\gamma^\prime)$.

To investigate the behavior of $\mu$ in the limit of vanishingly small strain rates, $\dot\gamma\to 0$, we can
write an alternative scaling form for $\mu$ by using Eq.~(\ref{epx}).  Again noting  that the critical exponents $y$ for $p$ and $\sigma$ are equal, we get,
\begin{equation}
\mu\equiv\dfrac{\sigma}{p}=\tilde h_{1\pm}\left(\dfrac{\dot\gamma}{|\delta\phi|^{z\nu}}\right)
+|\delta\phi|^{\omega\nu}\tilde h_{2\pm}\left(\dfrac{\dot\gamma}{|\delta\phi|^{z\nu}}\right),
\label{emuscale}
\end{equation}
where $\pm$ indicate the scaling functions above and below $\phi_J$ respectively.
Thus as $\dot\gamma\to 0$, we expect the limiting behavior,
\begin{equation}
\mu=\tilde h_{1\pm}(0)+|\delta\phi|^{\omega\nu}\tilde h_{2\pm}(0).
\label{emuscale2}
\end{equation}
Taking the limit of Eq.~(\ref{emuscale2}) as $\phi\to\phi_J$ from below we therefore get $\mu(\phi_J^-)=\tilde h_{1-}(0)$, while taking the limit $\phi\to\phi_J$ from above we get $\mu(\phi_J^+)=\tilde h_{1+}(0)$.  Since there is no reason why we should have $\tilde h_{1-}(0)=\tilde h_{1+}(0)$, Eq.~(\ref{emuscale2}) implies that as $\dot\gamma\to 0$, $\mu$ takes a discontinuous jump $\Delta\mu=\tilde h_{1+}(0)-\tilde h_{1-}(0)$ at $\phi_J$.  Looking at our data in Fig.~\ref{sop}, however, we cannot detect any suggestion of such a discontinuity in $\mu$ at $\phi_J$; the expected discontinuity may be too small, or may not become sharp enough until even smaller $\dot\gamma$ is reached.

We do, however, see what appears to be a discontinuous slope in $\mu$ at $\phi_J$, as $\dot\gamma\to 0$.  This is also a consequence of Eq.~(\ref{emuscale2}).  If we assume that $\omega\nu=1$, then as $\phi\to\phi_J$ from below we have $d\mu/d\phi=-\tilde h_{2-}(0)$, while for $\phi\to\phi_J$ from above we have $d\mu/d\phi = +\tilde h_{2+}(0)$, giving a discontinuity in the slope $\Delta (d\mu/d\phi) = \tilde h_{2+}(0)+\tilde h_{2-}(0)$.
The physical reason for this discontinuous slope is straightforward: As $\dot\gamma\to 0$ below $\phi_J$, $\mu$ is the ratio of Bagnold coefficients, $\mu=B_\sigma/B_p$, as $p$ and $\sigma$ each individually goes to zero; above $\phi_J$, $\mu$ is the ratio of the shear and pressure components of the yield stress, $\mu=\sigma_0/p_0$.  There is no reason that the $\phi$ dependence of $B_\sigma/B_p$ should be smoothly related to the $\phi$ dependence of $\sigma_0/p_0$, and this is formalized in the scaling of Eq.~(\ref{emuscale2}).

\subsection{Hard-core limit}

Our scaling analysis in the previous sections required us to consider corrections-to-scaling in order to arrive at consistent results.  It is therefore puzzling how Peyneau and Roux \cite{Roux} managed to get  from the constituent equations (\ref{ec1}) and (\ref{ec2}) the same exponents as we find here, {\em without} having to consider corrections-to-scaling.  Although they simulate with soft-core particles as we do, they claim that their particles are sufficiently stiff (and $\dot\gamma$ sufficiently small) that their results were all obtained in the hard-core limit where the inertial number $I\sim \dot\gamma/\sqrt{p}$ is independent of the specific values of $p$ and $\dot\gamma$, and only depends on the packing fraction $\phi$, as discussed in Sec.~\ref{sConstituent}.

From Eq.~(\ref{ecorrection}) we see that $p$ (and similarly $\sigma$ and so $\mu$) depends on the scaling variable $x\equiv \delta\phi/\dot\gamma^{1/z\nu}$.  Since the hard-core limit is characterized by sufficiently small $\dot\gamma$, where $|x|$ is therefore large, the crossover from the hard-core to the soft-core region is set by the scaling function to be at some particular value $x^*$;  $|x|\gg x^*$ is the hard-core region while $|x|\le x^*$ is the soft-core region.  Equivalently, if $\dot\gamma^*(\phi)\equiv  |\delta\phi/x^*|^{z\nu}$, then $\dot\gamma\ll \gamma^*$ is the hard-core region.  Since the data we have used in our fits all satisfy $|x|\le x_\mathrm{max}$, for some suitably small $x_\mathrm{max}$, our scaling analysis above has used data that are all explicitly in the soft-core region. 
It is therefore of interest to instead consider our data that are in the hard-core region, and see what exponents are obtained from an analysis of those results.

\begin{figure}[h!]
\includegraphics[width=3.2in]{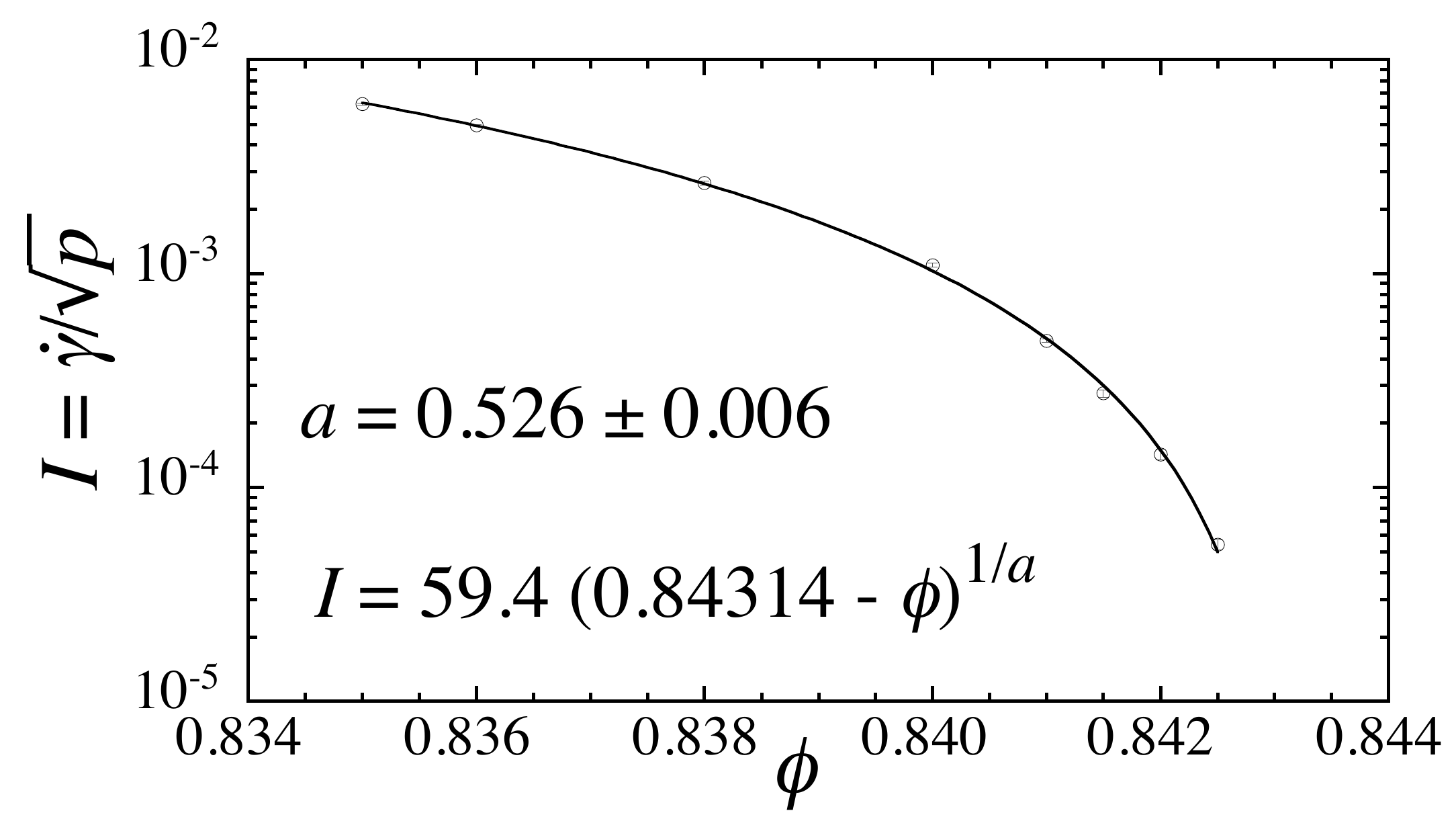}
\caption{Inertial number $I=\dot\gamma/\sqrt{p}$ vs packing fraction $\phi$, for our data that are in the hard-core limit, $\phi<\phi_J$ and $\dot\gamma\to 0$.
}
\label{I-vs-phi}
\end{figure}

From Fig.~\ref{Bp-Bs}a for $B_p=p/\dot\gamma^2$, we see that we have data that are in the hard-core limit, with $B_p$ independent of the strain rate  $\dot\gamma$ at sufficiently small $\dot\gamma$, for packing fractions up to the value $\phi=0.8425$.  Thus we are able to get hard-core results to within 0.1\% of $\phi_J\approx 0.84335$.  In Fig.~\ref{I-vs-phi}a we plot the inertial number $I=\dot\gamma/\sqrt{p}$ vs $\phi$ for our data points that are in the hard-core limit.  We see that we get down to a smallest value of $I_\mathrm{min}\approx 5\times 10^{-5}$.  In comparison, Peyneau and Roux \cite{Roux} consider two different numerical systems, with different particle stiffnesses, one of which extends down to $I_\mathrm{min}=10^{-5}$ and the other to $I_\mathrm{min}=3.2\times 10^{-5}$.   Fitting to values of $I_\mathrm{min}\le I\le 10^{-2}$, they find
for their two cases with different $I_\mathrm{min}$ the dilatancy exponents $a=0.42\pm0.02$ and $a=0.39\pm 0.01$ respectively.

Inverting the constituent equation (\ref{ec1}) to write $I\propto(\phi_J-\phi)^{1/a}$, we fit our hard-core data in Fig.~\ref{I-vs-phi} to this form, with $\phi_J$, $a$, and the proportionality constant as free fitting parameters.  We use the same range $I_\mathrm{min}\le I\le 10^{-2}$ as Peyneau and Roux \cite{Roux}.  We find $\phi_J=0.84314$, slightly smaller than the value 0.84335 obtained from our earlier analysis of data in the soft-core region.  We find a dilatancy exponent $a=0.526\pm 0.006$, larger than the value $a\approx 0.4$ found by Peyneau and Roux, and giving a value of $\beta=2/a\approx3.8$ that is significantly smaller than the $\beta=5.0\pm 0.4$ found from our earlier analysis, but is roughly equal to the value found in Fig.~\ref{fitparams2} provided we included a broad range of strain rates with $\dot\gamma_\mathrm{max}\approx 10^{-4}$.  
We thus conclude that, even in the hard-core region, our data do not get sufficiently close to the critical point that we can avoid the need for corrections-to-scaling.

We thus remain with the unanswered question as to why Peyneau and Roux \cite{Roux} found seemingly correct results without considering corrections-to-scaling.  
It is possible that this agreement is just fortuitous.  Although they claim that their results are in the hard-core limit, when they compare data for two different particle stiffnesses, their results for $\phi$ vs $I$ show a very small but noticeable and systematic difference at the smallest values of $I$ (see their Fig.~7), thus suggesting that the soft-core is influencing their results at the points closest to jamming ($I\to0$).  They also have a small, but measurable, finite size effect in their data (see their Figs.~6 and 8).
However their simulations differ from ours in several other ways.  They simulate at constant normal pressure, rather than constant volume.  It is claimed that fluctuation  and finite-size effects are reduced in the constant pressure ensemble.  However, even if so, our data are certainly accurate enough, and our system size ($N=262144$, compared to Peyneau and Roux's 4000) is certainly large enough, that this cannot be the source of the difference.  Peyneau and Roux \cite{Roux} simulate in three dimensions, while we are in two dimensions.  They use a monodisperse system, while we use a bidisperse system.  It thus may be that the magnitude of the corrections-to-scaling are affected by the dimensionality or dispersity of the system.

\begin{figure}[h!]
\includegraphics[width=3.2in]{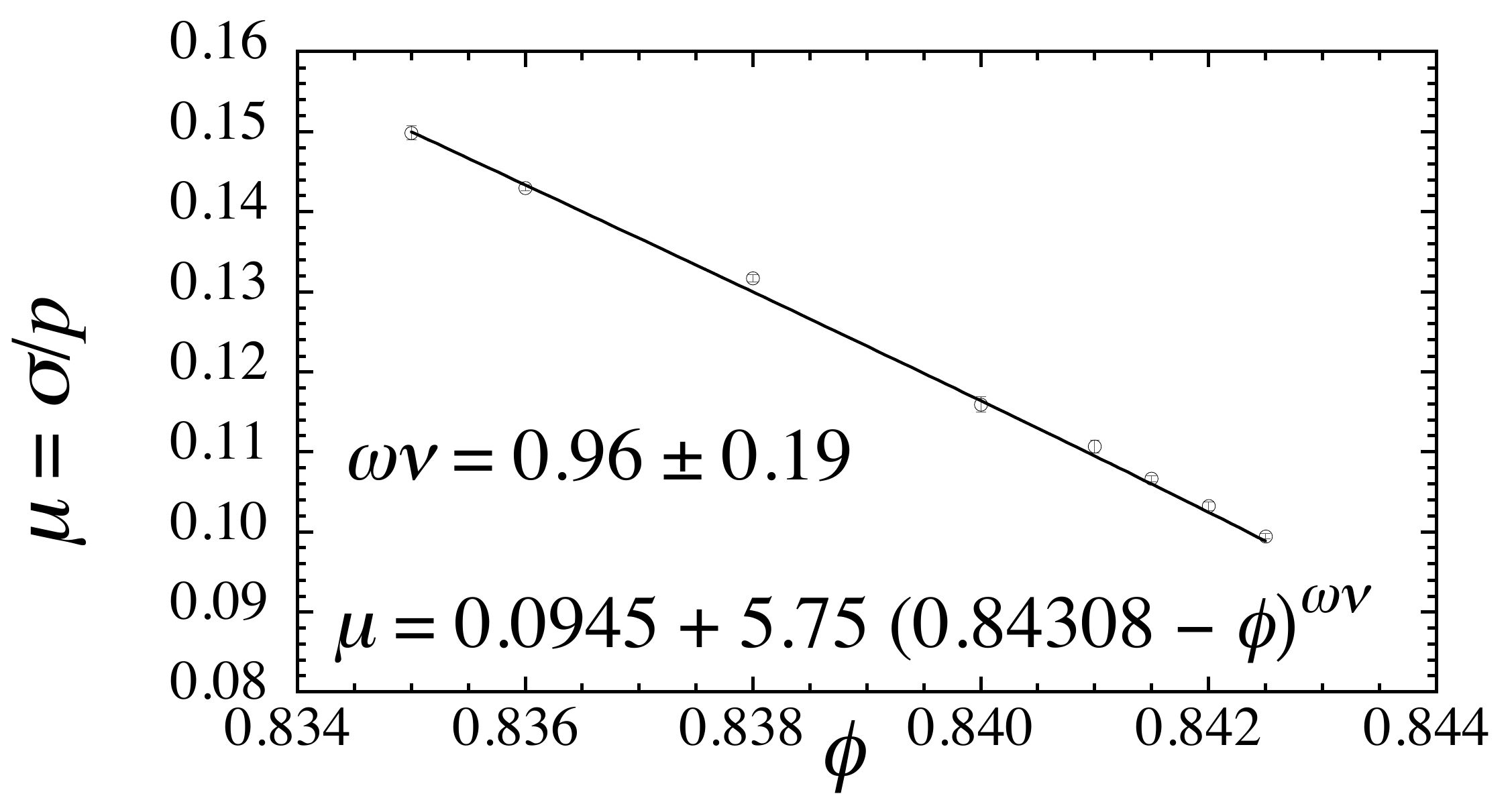}
\caption{Macroscopic friction $\mu\equiv \sigma/p$ vs packing fraction $\phi$, for our data that are in the hard-core limit, $\phi<\phi_J$ and $\dot\gamma\to 0$.
}
\label{mu-vs-phi}
\end{figure}

Finally, we consider the macroscopic friction $\mu$ for our data in the hard-core region.  Since our control parameter is $\phi$ rather than $p$, 
in Fig.~\ref{mu-vs-phi} we plot $\mu$ vs $\phi$ (rather than $I$), for the same hard-core data points as in Fig.~\ref{I-vs-phi}.  Fitting our data to Eq.~(\ref{emuphi}), we find $\phi_J\approx 0.84308$ and the exponent $\omega\nu\approx 0.96\pm 0.19$.  Thus, as in the analysis of Fig.~\ref{I-vs-phi}, the value of $\phi_J$ found here is somewhat smaller than found in our earlier analysis, but the value of $\omega\nu$ is in good agreement.

\section{Conclusions}
\label{sConclusion}

We have carried out constant volume simulations of a well studied model of frictionless disks in two dimensions that displays Bagnoldian rheology.  Simulating at shear strain rates $\dot\gamma$ slower than studied previously, we analyze our results for pressure $p$ and shear stress $\sigma$ according to a critical scaling ansatz.  We show that, for the range of parameters considered here, a simple scaling analysis fails to give consistent results as we vary the window of data about the jamming transition that is used to fit to the scaling expression; parameter values are found to systematically vary with the width of the window of data used.  Our results highlight that, in carrying out a scaling analysis of critical parameters, it is not sufficient to do a fit to the data and find a good looking scaling collapse, as in our Fig.~\ref{p-s-scaled}; rather it is essential to check the stability of the fitted critical parameters to a narrowing of the window of data about the critical point, as shown in our Figs.~\ref{fitparams} and \ref{fitparams2}.

We show, however, that consistent results are found once we include corrections-to-scaling in the analysis.  The exponent $\beta$ that describes the divergence of the hard-core Bagnold coefficients $B_p$ and $B_\sigma$ is found to be noticeably larger than the value $\beta=4$ predicted by the theory of Otsuki and Hayakawa \cite{OH1,OH2}.  Our value $\beta\approx 5.0\pm 0.4$ is consistent with earlier numerical simulations by Peyneau and Roux \cite{Roux} who found $\beta\approx 5.0\pm 0.3$, and is closer to the value $\beta\approx 5.7$ predicted theoretically  by the recent work of DeGiuli et al. \cite{DeGiuli}.  
Our results therefore cast significant doubt on the mean-field calculations of Otsuki and Hayakawa \cite{OH1,OH2} while lending support to the theoretical arguments of DeGiuli et al. \cite{DeGiuli}.

We have considered the macroscopic friction $\mu$, and shown how the dependence of $\mu$ on $\phi$ is directly related to corrections-to-scaling.  While we have found it difficult to determine an accurate value of the relevant correction-to-scaling exponent $\omega\nu$, our results are consistent with the value $\omega\nu\approx 1$, in agreement with the claims of DeGiuli et al. \cite{DeGiuli} and consistent with the numerical results of Peyneau and Roux \cite{Roux}. 

Our detailed comparisons with the earlier simulations of Peyneau and Roux \cite{Roux} suggest that the magnitude of the corrections-to-scaling may be affected by the dimensionality of the system, or the size dispersity of the particles.  This remains for further investigation.

\section*{Acknowledgements}

This work was supported by National Science Foundation Grant No. DMR-1205800, Swedish Research Council Grant No. 2010-3725, and the Dutch Organization for Scientific Research (NWO). Simulations were performed on resources provided by the Swedish National Infrastructure for Computing (SNIC) at PDC and HPC2N.  We wish to thank M. Wyart for helpful discussions. 

\section*{Appendix A}

In this appendix we demonstrate that our system with $N=262144$ total particles is big enough so that there are no finite size effects in our data, and we comment on the applicability of a finite-size-scaling approach to analyze our system.  For a continuous phase transition there is usually a correlation length $\xi$ that diverges as the critical point is approached.  When $\xi$ becomes comparable to, or bigger than, the system length $L$, finite size effects become manifest.  If we wish to do critical scaling in the infinite system size limit, such as we have done in this work, we therefore need to make certain that our system size is sufficiently large that $L\gg\xi$ for all the parameters $(\phi,\dot\gamma)$ where we carry out our simulations.  While in the present model it is not straightforward to measure $\xi$ directly, we can nevertheless check that we are in the appropriate limit by comparing results from simulations of different system sizes $L$.

\begin{figure}[h!]
\includegraphics[width=3.2in]{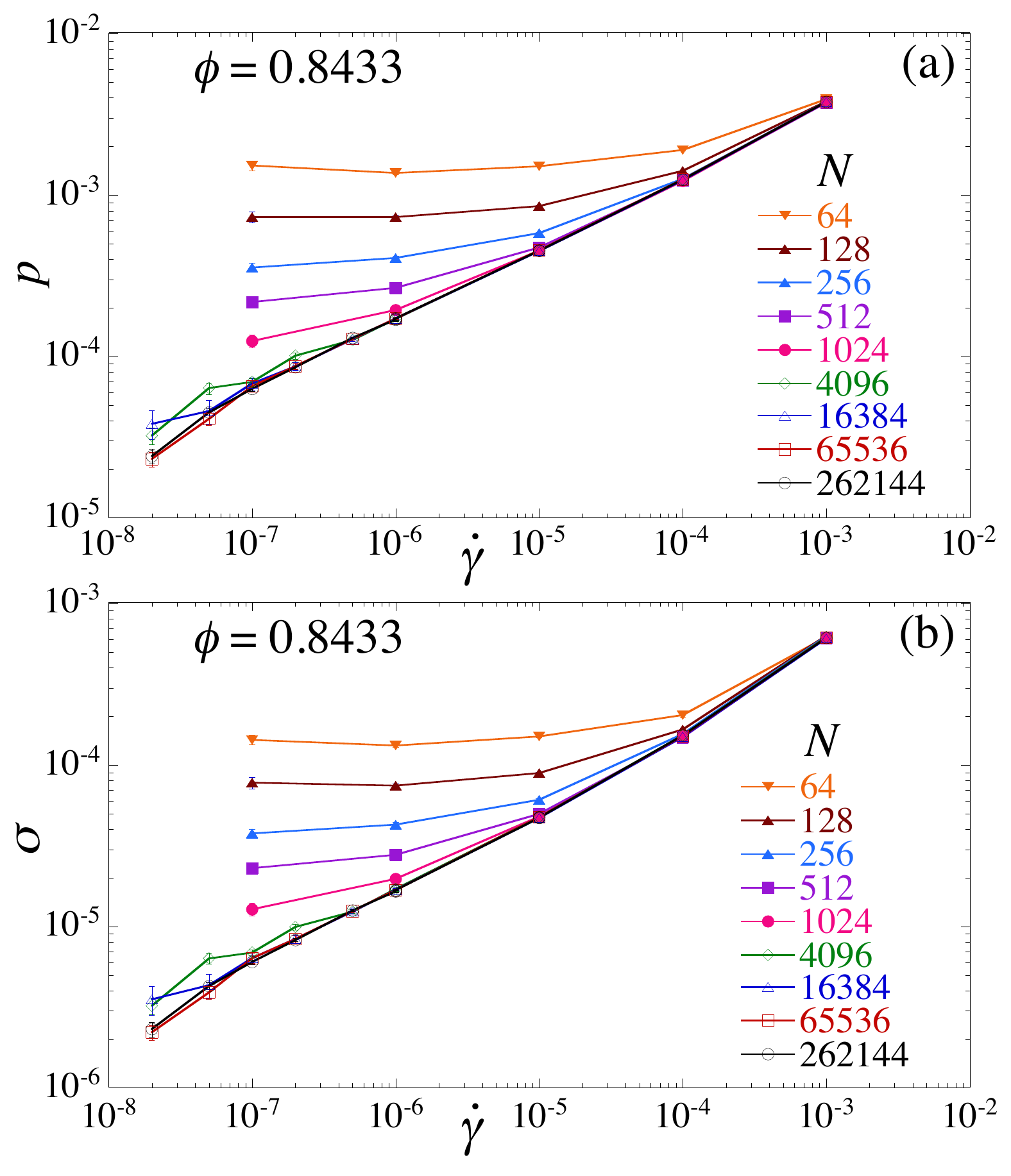}
\caption{(Color online)  (a) Pressure $p$ and (b) shear stress $\sigma$ vs shear strain rate $\dot\gamma$ for systems with different number of particles $N$, at the packing fraction $\phi=0.8433$, which is just slightly below the jamming $\phi_J$.
}
\label{finitesize}
\end{figure}

Since $\xi$ should diverge at $\phi=\phi_J$ as $\dot\gamma\to 0$, for the parameters we simulate, the correlation length $\xi$ will be largest at our smallest $\dot\gamma$ at the $\phi$ that is closest to $\phi_J$.  It thus suffices to look at the behavior of our system, as a function of $\dot\gamma$ and particle number $N$, close to $\phi_J$. In Fig.~\ref{finitesize} we therefore plot the  pressure $p$ and shear stress $\sigma$ vs shear strain rate $\dot\gamma$, for several different system sizes as measured by the number of particles $N$.  Our results are for the packing fraction $\phi=0.8433$, which our scaling analysis indicates is just very slightly below the jamming $\phi_J$.  

For small $N$ we see that $p$ and $\sigma$ plateau to  constant values as $\dot\gamma$ decreases; this plateau is a consequence of $\xi$ becoming comparable to the system length $L\sim N^{1/d}$, in $d$ dimensions.  The value of $\dot\gamma^*$ at which the plateau sets in, and the values $p^*$ and $\sigma^*$ on the plateau, decrease as $N$ increases, since as $N$ increases we are able to get closer to the critical point before the condition $L\sim\xi$ sets in.  For the largest system sizes however, we see no such plateau, indicating that $L\gg\xi$ even at the smallest $\dot\gamma$.  Comparing the two largest systems sizes $N=65536$ and $262144$, we see no dependence of our data on $N$, within the estimated statistical error.  This confirms that our system with $N=262144$ particles is safely in the infinite size limit for the parameters where we simulate, and that any finite size effects are completely negligible, thus justifying our use of the scaling of Sec.~\ref{sResults}.  This is the main point of this appendix.

We may note that it is sometimes possible to determine critical exponents, characteristic of the infinite size limit, by exploiting the dependence of quantities on system size.  This method, known as finite-size-scaling, is based on viewing the inverse of the system length $L^{-1}$ as a new control parameter that vanishes at the critical point, and generalizing the scaling of Eq.~(\ref{escale0}) to the form,
\begin{equation}
pb^{y/\nu}=f(\delta\phi b^{1/\nu}, \dot\gamma b^z, w_1 b^{-\omega_1}, w_2 b^{-\omega_2},\dots,L^{-1}b).
\label{escaleL}
\end{equation}
Choosing $b=L$, and keeping only the leading irrelevant scaling variable, then gives,
\begin{equation}
pL^{y/\nu}=f(\delta\phi L^{1/\nu}, \dot\gamma L^z, w L^{-\omega}, 1).
\label{escaleL2}
\end{equation}
The scaling equation above is more complicated than what we have considered previously, i.e. Eq.~(\ref{escale1}).   Even if we regard the leading irrelevant scaling variable as small and negligible, $w\approx 0$, the right hand side of Eq.~(\ref{escaleL2}) still involves two independent scaling variables, $\delta\phi L^{1/\nu}$ and $\dot\gamma L^z$.  Since we do not a priori know either of the scaling exponents $1/\nu$ or $z$, nor the value of $\phi_J$, proceeding with Eq.~(\ref{escaleL2}) would require us to explore a three dimensional parameter space $(\phi, \dot\gamma, L)$ rather than the two dimensional parameter space $(\phi,\dot\gamma)$ considered in Sec.~\ref{sResults}.  To simplify, we need to eliminate one of the control parameters $\phi$ or $\dot\gamma$, so as to reduce the problem to a single scaling variable.

If we can do quasistatic shearing simulations \cite{OTFSS}, with $\dot\gamma\to 0$, one can then write (assuming $w=0$),
\begin{equation}
pL^{y/\nu}=f(\delta\phi L^{1/\nu}, 0, 0, 1).
\label{escaleL4}
\end{equation}
Plotting $pL^{y/\nu}$ vs $(\phi-\phi_J)\phi L^{1/\nu}$, and requiring the data to collapse to a common curve for different $L$, then determines the exponents $y/\nu$ and $1/\nu$, as well as $\phi_J$.  However our simulations in the present work are all at finite $\dot\gamma$, so this approach is not possible for us.

If we knew the exact location of the jamming point $\phi_J$, we could then simulate at $\phi=\phi_J$, and write (assuming $w=0$) \cite{Hatano4},
\begin{equation}
pL^{y/\nu}=f(0,\dot\gamma L^z,0,1).
\label{escaleL3}
\end{equation}
Requiring $p$ to become independent of $L$ as $L\to\infty$ then requires that $f(0,x,0,1)\sim x^{y/z\nu}$ as $x\to\infty$, thus giving in the infinite size limit the critical rheology at $\phi_J$, $p\sim\dot\gamma^q$ with $q=y/z\nu$ in agreement with Eq.~(\ref{eq1}).
In the opposite limit of $x\to 0$, assuming $f(0,x,0,1)\to\mathrm{constant}$ gives, $\lim_{\dot\gamma\to 0} p \equiv p^*\sim L^{-y/\nu}$, and the crossover to this low strain rate limit occurs at $\dot\gamma^*\sim L^{-z}$.  Both $p^*$ and $\dot\gamma^*$ thus scale to zero as $L$ increases, in qualitative agreement with what we see in Fig.~\ref{finitesize}.
Plotting $pL^{y/\nu}$ vs $\dot\gamma L^z$, and requiring the data to collapse to a common curve for different $L$, then determines the exponents $y/\nu$ and $z$.  Hwever, even if we could do this, it does not allow us to determine the critical exponent $\beta$ of the transport coefficient, which is the focus of the present work.  From Eq.~(\ref{ebeta1}) we have for Bagnold rheology ($n=2$), $\beta=2z\nu-y=\nu(2z-y/\nu)$.  The finite-size-scaling method of Eq.~(\ref{escaleL3})  determines $z$ and $y/\nu$, but does not determine $\nu$, thus preventing us from determining $\beta$.
Of course our ability to even attempt the above analysis depends on our knowing the exact value of $\phi_J$, which we know only approximately, and only because we have already done the infinite size scaling analysis of Sec.~\ref{sResults}.

Finally, we note that since our scaling analysis (Sec.~\ref{sResults}) in the infinite system size limit indicated that corrections-to-scaling from the leading irrelevant variable are important for the strain rates $\dot\gamma$ studied here, one should expect such corrections to be important in systems of finite size as well.
If so, our simple Eq.~(\ref{escaleL3}) is not sufficient to describe our finite size data, but rather we should expand Eq.~(\ref{escaleL2}) for small $w$, and then set $\delta\phi=0$,  to obtain,

\begin{equation}
pL^{y/\nu}=f_1(\dot\gamma L^z) + L^{-\omega}f_2(\dot\gamma L^z).
\label{escaleL5}
\end{equation}
Indeed, we have found such corrections-to-scaling to be important in a finite-size-scaling analysis of a related model with Newtonian rheology, both for scaling with $\delta\phi$ in the quasistatic limit $\dot\gamma\to 0$ \cite{OTFSS}, and for scaling with $\dot\gamma$ at $\phi=\phi_J$ \cite{noteSM}.

To summarize, a finite-size-scaling analysis for our model is problematic for many reasons: (i) To determine all desired critical exponents we would need to deal with the scaling equation~(\ref{escaleL2}) which involves two independent scaling variables and a three dimensional parameter space $(\phi,\dot\gamma,L)$, for which there is no simple way forward; (ii) we cannot simplify to the single variable scaling with $\delta\phi$, as in Eq.~(\ref{escaleL4}), as we are not in the quasistatic limit; (iii) we cannot simplify to the single variable scaling with $\dot\gamma$, as in Eq.~(\ref{escaleL3}), as we do not a priori know the exact value of $\phi_J$; (iv) even if we could attempt scaling as in Eq.~(\ref{escaleL3}), our analysis would be complicated by corrections-to-scaling; (v) and finally, even if we could successfully carry out a finite-size-scaling analysis based on Eq.~(\ref{escaleL5}), that analysis would still not be sufficient to allow us to determine the value of the transport coefficient exponent $\beta$.
We therefore have chosen not to pursue a detailed finite-size-scaling analysis for the present model, but rather to focus our scaling analysis on behavior in the infinite size limit.

\section*{Appendix B}

The observation that critical exponents for the {\em static}, compression-driven, jamming transition appear to be the same in two as in three dimensions \cite{OHern} has lead to the speculation that $d=2$ may be at or above the upper critical dimension $d_{uc}$ for the jamming transition. An analysis by Wyart et al. \cite{Wyart2}, considering the spatial fluctuations of the contact number, argued that $d_{uc}=2$. Further evidence  that $d_{uc}\le 2$ was claimed from a finite-size scaling analysis of contact number vs pressure in numerical simulations by Goodrich et al. \cite{Goodrich1}.  Exactly at $d_{uc}$, scaling variables acquire multiplicative logarithmic corrections \cite{Wegner}.  Evidence for such logarithmic corrections was claimed in finite-size-scaling analyses of contact number vs pressure \cite{Goodrich2} and shear strain vs pressure \cite{Deen}, in mechanically stable packings of two dimensional frictionless disks compressed above the static jamming transition.

Although the discussion and evidence that $d_{uc}=2$ for the jamming transition have pertained {\em only} to the behavior of soft-core disks isotropically (on average) compressed above the {\em static} jamming transition, one can wonder if $d_{uc}=2$ may hold as well for the {\em dynamic} shear-driven jamming transition considered in this work.  In such a case, the logarithmic corrections change the scaling of Eq.~(\ref{escale0}) to the form \cite{Lubeck1},
\begin{equation}
pb^{y/\nu}|\ln b|^{c_p}=f(\delta\phi b^{1/\nu}|\ln b|^{c_\phi}, \dot\gamma b^z |\ln b|^{c_{\dot\gamma}}),
\label{eA1}
\end{equation}
where the leading algebraic exponents $y/\nu$, $1/\nu$ and $z$ take their mean-field values, and the new logarithmic exponents are $c_p$, $c_\phi$ and $c_{\dot\gamma}$; we have ignored for simplicity the irrelevant variables $w_i$.  One may now choose the length rescaling factor $b$ so that $\dot\gamma b^z |\ln b|^{c_{\dot\gamma}}=1$.  To leading order as $\dot\gamma\to 0$, this results in \cite{Lubeck2},
\begin{equation}
p=\dot\gamma^{y/z\nu}|\ln \dot\gamma|^{c_1}\tilde f\left(\dfrac{\delta\phi}{\dot\gamma^{1/z\nu}|\ln \dot\gamma|^{c_2}}\right),
\label{eA2}
\end{equation}
which is the analog of Eq.~(\ref{escale1}).  Exactly at jamming, $\delta\phi=0$, and the rheology at criticality becomes,
\begin{equation}
p\sim \dot\gamma^{q}|\ln \dot\gamma|^{c_1},\quad\mathrm{at}\quad \phi=\phi_J,
\label{eA3}
\end{equation}
with $q=y/z\nu$ as before.
Alternatively, one can choose $b$ so that $|\delta\phi| b^{1/\nu}|\ln b|^{c_\phi}=1$, in which case to leading order as $\delta\phi\to 0$ one gets below jamming,
\begin{equation}
p=|\delta\phi|^y\big|\ln|\delta\phi|\big|^{\tilde c_1}\tilde g\left(\dfrac{\dot\gamma}{|\delta\phi|^{z\nu}\big|\ln|\delta\phi|\big|^{\tilde c_2}}\right).
\end{equation}
Since below jamming we expect $p\sim\dot\gamma^2$, we then have for the scaling of the Bagnold coefficient in the hard-core $\dot\gamma\to 0$ limit,
\begin{equation}
p/\dot\gamma^2\sim |\delta\phi|^{-\beta}\big|\ln|\delta\phi|\big|^c,\quad\mathrm{for}\quad \dot\gamma\to 0,\quad\phi<\phi_J,
\label{eA4}
\end{equation}
with $\beta=2z\nu-y$ as before.

We would now like to test our numerical results for evidence of such logarithmic corrections to scaling.  In particular we wish to see if such logarithmic corrections could give a self-consistent explanation for our results in Sec.~\ref{sResults},  without having to introduce the correction-to-scaling term from the leading irrelevant variable, as done in Sec.~\ref{ssCorrections}.  

However, there are many difficulties with attempting to fit to either Eqs.~(\ref{eA2}), (\ref{eA3}) or (\ref{eA4}).  We cannot use Eq.~(\ref{eA3}) directly, since we do not a priori know the value of $\phi_J$; using an incorrect value of $\phi$ slightly off from $\phi_J$ would skew data at the smallest $\dot\gamma$ away from the form of Eq.~(\ref{eA2}) and so a fit to Eq.~(\ref{eA2}) would give spurious results. It is difficult to use Eq.~(\ref{eA4}) since our simulations are not explicitly in the hard-core $\dot\gamma\to 0$ limit; the $\dot\gamma$ dependence of $p/\dot\gamma^2$ sets in at ever decreasing values of $\dot\gamma$ as one gets closer to $\phi_J$.  Moreover,
it can be  exceedingly difficult to numerically distinguish the form $x^b|\ln x|^c$ from the form $x^{b^\prime}$ when the range of data for $x$ is limited, as it is in our case.  The success of such fits generally depends on knowing in advance the mean-field value of the leading exponent $b$, and often the exact location of the critical point.

For example, in Ref.~\cite{Goodrich2} the authors do a finite-size scaling analysis of the average contact number $Z$ with pressure $p$ and system size $N$.  Using a scaling form similar to our Eq.~(\ref{eA2}) with $Z-Z_c$ playing the role of our $p$, $p$ playing the role of our $\delta\phi$, and $1/N$ playing the role of our $\dot\gamma$,  they fit to the form,
\begin{equation}
Z-Z_c^N=\frac{1}{N}f\left(\dfrac{pN^2}{|\ln N|^{c_2}}\right).
\end{equation}
However, in their case they know the exact location of their critical point, $p=0$ and $Z_c^N=2d-2d/N_0$ the isostatic value for a system with $N_0$ non-rattler particles \cite{Goodrich2}.  Futhermore, the mean-field exponents relevant to this situation are believed to be known, and these values are used in their fits, i.e. the analogs of $y/z\nu$ and $1/z\nu$ in Eq.~(\ref{eA2}) are here 1 and 2 respectively.  Moreover, the authors assume, with no justification given, that the analog of the exponent $c_1$ in Eq.~(\ref{eA2}) vanishes, and hence there is no logarithmic correction to the scaling of the contact number $z$.  Thus only the single exponent $c_2$ is to be determined from the fit, and still the authors never show any quantitative measure of the success of their fit (such as the $\chi^2_\mathrm{dof}$) or test the stability of their obtained value of the exponent $c_2$ to changes in the window of data used in the fit.

\begin{figure}[h!]
\includegraphics[width=3.2in]{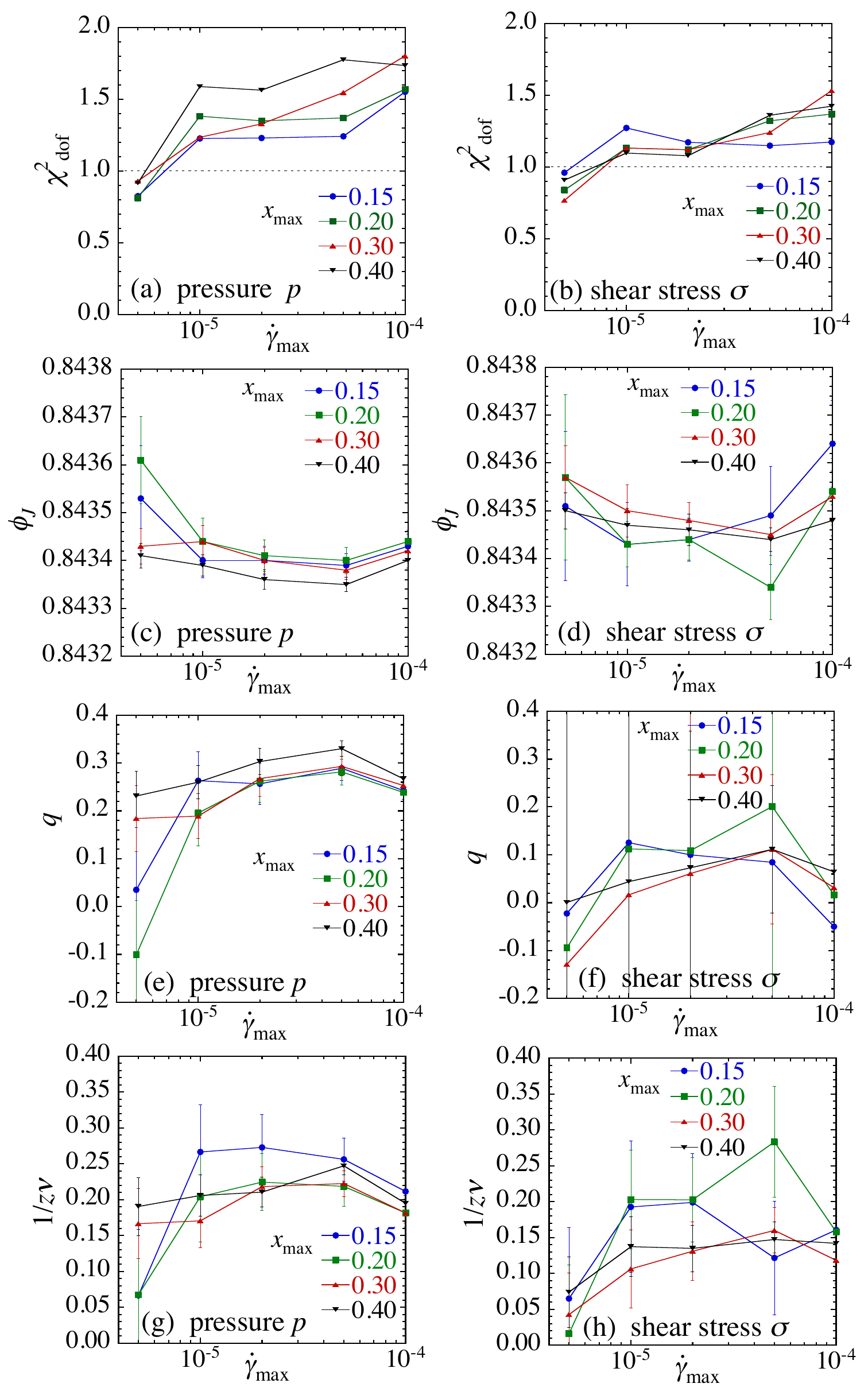}
\caption{(Color online)  Fits of pressure $p$ (left column) and shear stress $\sigma$ (right column) to the scaling form of Eq.~(\ref{eA2}) including presumed logarithmic corrections to scaling.  Different curves represent different cutoffs $x_\mathrm{max}$ on the scaling variable; results are plotted vs the cutoff on the strain rate $\dot\gamma_\mathrm{max}$.  We show results for the chi squared per degree of freedom of the fit $\chi^2_\mathrm{dof}$, the jamming fraction $\phi_J$, and the exponents $q$ and $1/z\nu$.
}
\label{fA1}
\end{figure}

\begin{figure}[h!]
\includegraphics[width=3.2in]{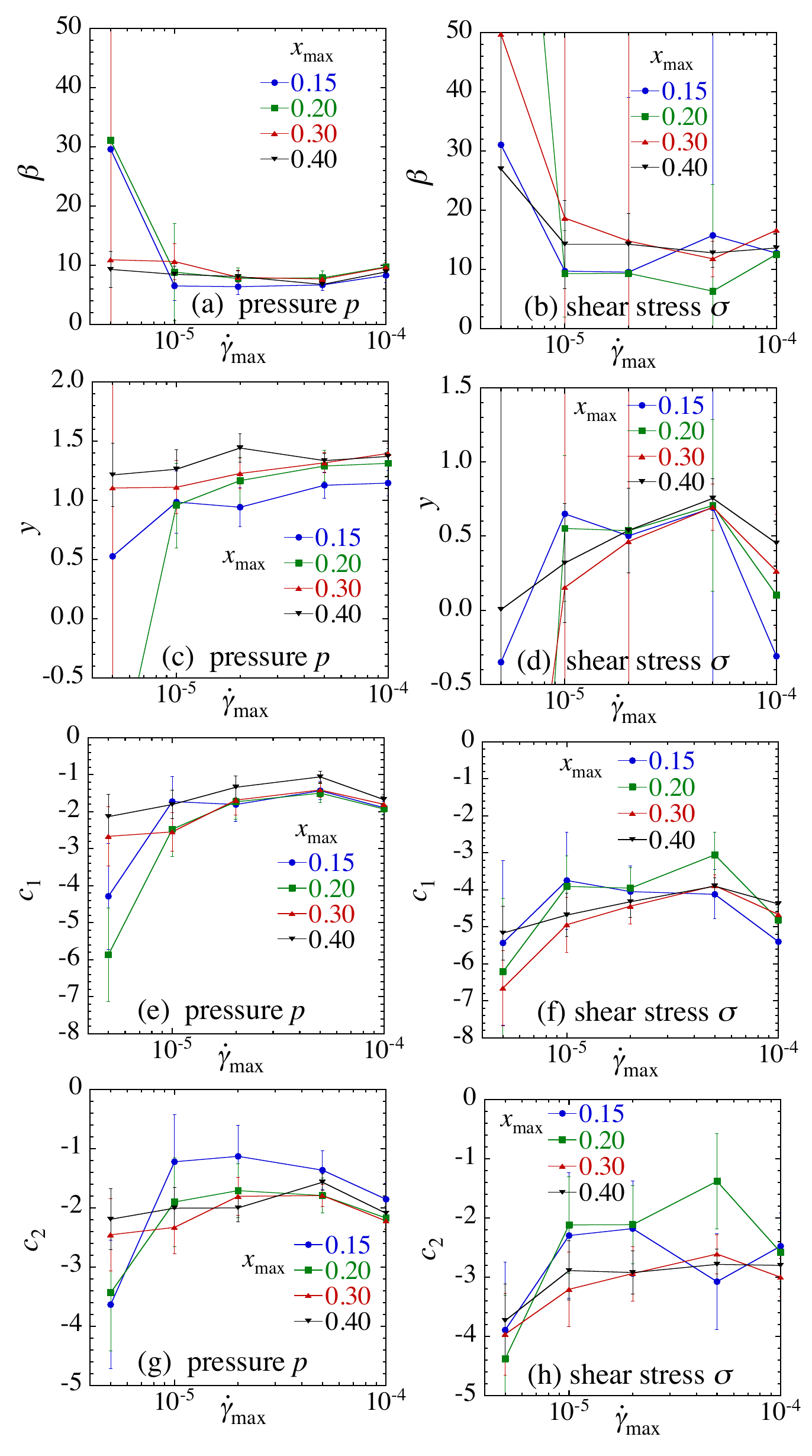}
\caption{(Color online)  Fits of pressure $p$ (left column) and shear stress $\sigma$ (right column) to the scaling form of Eq.~(\ref{eA2}) including presumed logarithmic corrections to scaling.  Different curves represent different cutoffs $x_\mathrm{max}$ on the scaling variable; results are plotted vs the cutoff on the strain rate $\dot\gamma_\mathrm{max}$.  We show results for the exponent $\beta=(2-q)z\nu$ and $y=qz\nu$, and the new exponents $c_1$ and $c_2$ associated with the logarithmic corrections.
}
\label{fA2}
\end{figure}

In contrast, for our Eq.~(\ref{eA2}) we do not know the precise value of $\phi_J$, nor the exponents $q=y/z\nu$ or $1/z\nu$, nor the new exponents $c_1$ and $c_2$; these are all quantities we wish to determine from the fit.  Nevertheless we can attempt to see how well our data fit the form of Eq.~(\ref{eA2}), where we approximate the scaling function $\tilde f(x)$ by the exponential of a fifth order polynomial as in Eq.~(\ref{escalfun}), and use the polynomial coefficients, $\phi_J$, $q=y/z\nu$, $1/z\nu$, $c_1$ and $c_2$ as free fitting parameters.  We compare the results of our fits varying the window of data used, $\dot\gamma\le\dot\gamma_\mathrm{max}$ and $|x|\le x_\mathrm{max}$, as we decrease the limiting values $\dot\gamma_\mathrm{max}$ and $x_\mathrm{max}$ just as we have done in the earlier Sec.~\ref{ssWithout}.  We use the following procedure: for given values of $\dot\gamma_\mathrm{max}$ and $x_\mathrm{max}$ we use as initial guesses for the fit parameters the values obtained from our earlier fits of Sec.~\ref{ssWithout} at the corresponding $\dot\gamma_\mathrm{max}$ and $x_\mathrm{max}$, together with $c_1=c_2=0$; using these parameters, we select the data to be used in the fit according to the criteria $\dot\gamma\le\dot\gamma_\mathrm{max}$ and $|x|=|\delta\phi/\dot\gamma^{1/z\nu}|\le x_\mathrm{max}$; we then carryout the fit letting all fitting parameters, including $c_1$ and $c_2$, vary.  Our results, independently fitting to both the pressure $p$ (left column) and the shear stress $\sigma$ (right column), are shown in Figs.~\ref{fA1} and \ref{fA2}.


Although we find that the quality of the fits, as measured by the $\chi^2_\mathrm{dof}$, are reasonably good (at least as good as in Fig.~\ref{chisq-p-s} for the fits ignoring the logarithmic corrections), nevertheless the outcomes of these fits cannot be taken as evidence for the correctness of the scaling assumption of Eq.~(\ref{eA2}).  If Eq.~(\ref{eA2}) were correct, we would expect to see the fitted parameters become independent of the cutoffs $\dot\gamma_\mathrm{max}$ and $x_\mathrm{max}$ as these cutoffs decreased.  However the values of the exponents $q$ and $1/z\nu$ (and correspondingly $\beta$ and $y$), as well as the new exponents $c_1$ and $c_2$, vary considerably with $\dot\gamma_\mathrm{max}$ (in some cases even changing sign).  Moreover, we would expect the critical exponents to be consistent comparing values for $p$ vs for $\sigma$, while here we see noticeable differences, particularly for $q$, $y$, $c_1$, and $c_2$.  
We conclude that these fits are not reliable.  We believe the main problem is that the functional form of Eq.~(\ref{eA2}) poorly constrains the fit parameters; in particular it is difficult to distinguish the difference between the forms $x^b|\ln x|^c$ and $x^{b^\prime}$ over our limited range of data.  One can decrease $b$ and increase $c$ in the first to get results that are hard to distinguish from a given $b^\prime$ in the second.  Indeed we see in our fits that as $q$ and $1/z\nu$ get smaller (as $\dot\gamma_\mathrm{max}$ decreases), the corresponding $|c_1|$ and $|c_2|$ get larger.  We believe that the same issue of poor constraint is behind the huge error bars we find on some of our data points.

To get more meaningful results it is necessary to better constrain the fits, for example by fixing the values of the leading exponents $q$ and $1/z\nu$ to their {\em mean-field} values.  However, the values of these exponents are {\em not} uncontroversially known, and determining them is what is the main objective of this work.  Nevertheless, we can fix them according to the predictions of competing theoretical models, and then see if our numerical results become consistent with these theoretical predictions once we include the logarithmic corrections to scaling.

\begin{figure}[h!]
\includegraphics[width=3.2in]{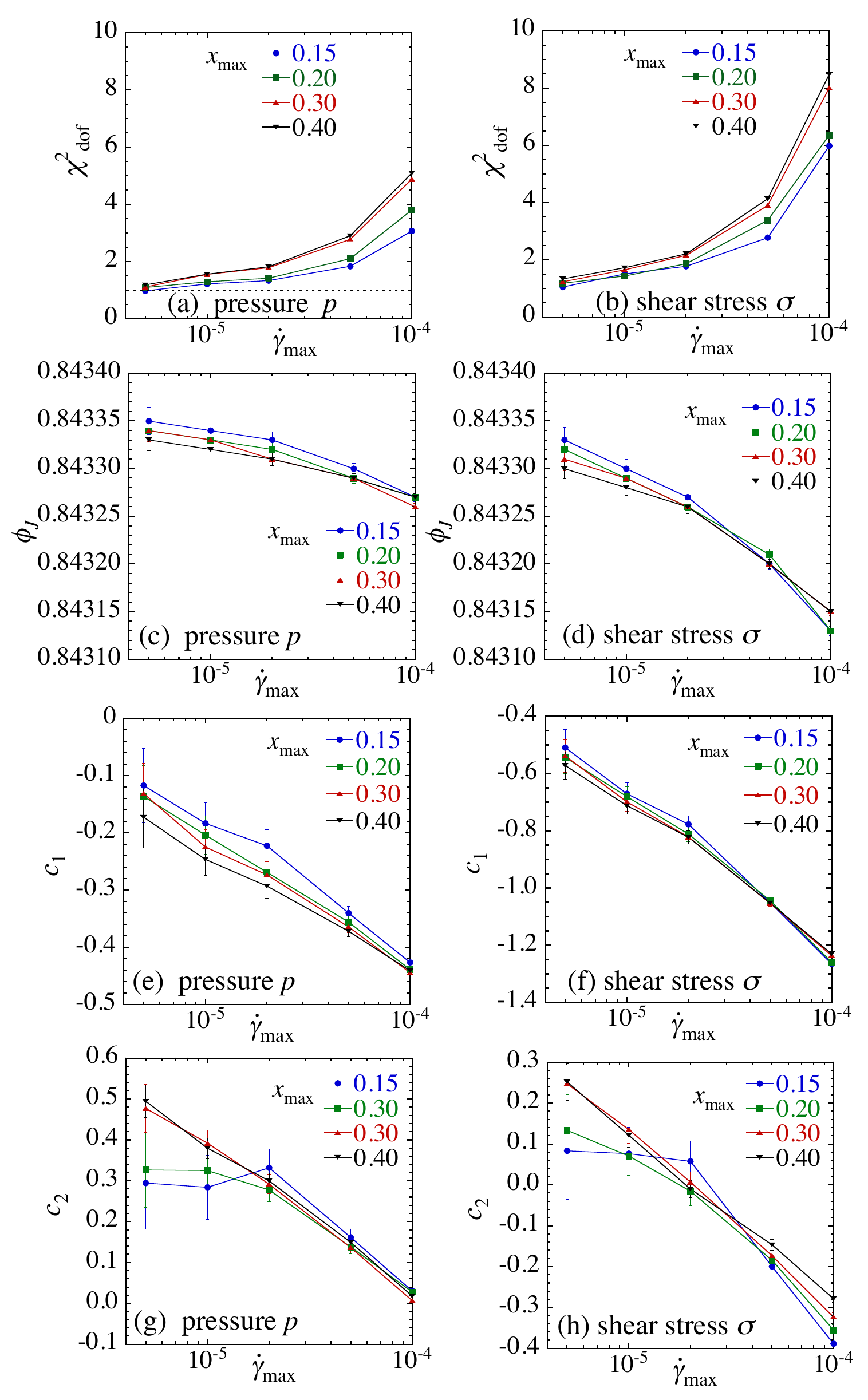}
\caption{(Color online)  Fits of pressure $p$ (left column) and shear stress $\sigma$ (right column) to the scaling form of Eq.~(\ref{eA2}) including presumed logarithmic corrections to scaling.  Here we fix the exponent values $q=1/z\nu=0.4$ as given by the theoretical prediction of Otsuki and Hayakawa \cite{OH1,OH2}. Different curves represent different cutoffs $x_\mathrm{max}$ on the scaling variable; results are plotted vs the cutoff on the strain rate $\dot\gamma_\mathrm{max}$.  We show results for the chi squared per degree of freedom $\chi^2_\mathrm{dof}$, the jamming fraction $\phi_J$, and the exponents $c_1$ and $c_2$ of the logarithmic corrections.
}
\label{fA3}
\end{figure}

We consider first the phenomenological mean-field theory of Otsuki and Hayakawa \cite{OH1,OH2} which gives $\beta=4$ and $y=1$; with these values, we have $1/z\nu=2/(\beta+y)=0.4$ and $q=2y/(\beta+y)=0.4$.  Fixing $q$ and $1/z\nu$ to these values we proceed as before, letting all other parameters vary in our fit.  In Fig.~\ref{fA3} we present the resulting values of $\chi^2_\mathrm{dof}$, $\phi_J$, $c_1$ and $c_2$, as functions of the cutoffs $\dot\gamma_\mathrm{max}$ and $x_\mathrm{max}$.  We show results from both fits to pressure (left column) and to shear stress (right column). We see that $c_1$ and $c_2$ continue to increase as $\dot\gamma_\mathrm{max}$ decreases, instead of saturating to a constant value, and moreover there is a significant difference between the values of $c_1$ and $c_2$ obtained from the fits to the pressure as compared with the values obtained from the shear stress.  Because of the clear dependence of the exponents $c_1$ and $c_2$ on the window of data used in the fit, we conclude that the logarithmic corrections of Eq.~(\ref{eA2}) do {\em not} lead to agreement between our results and the predictions of Otsuki and Hayakawa \cite{OH1,OH2}.

\begin{figure}[h!]
\includegraphics[width=3.2in]{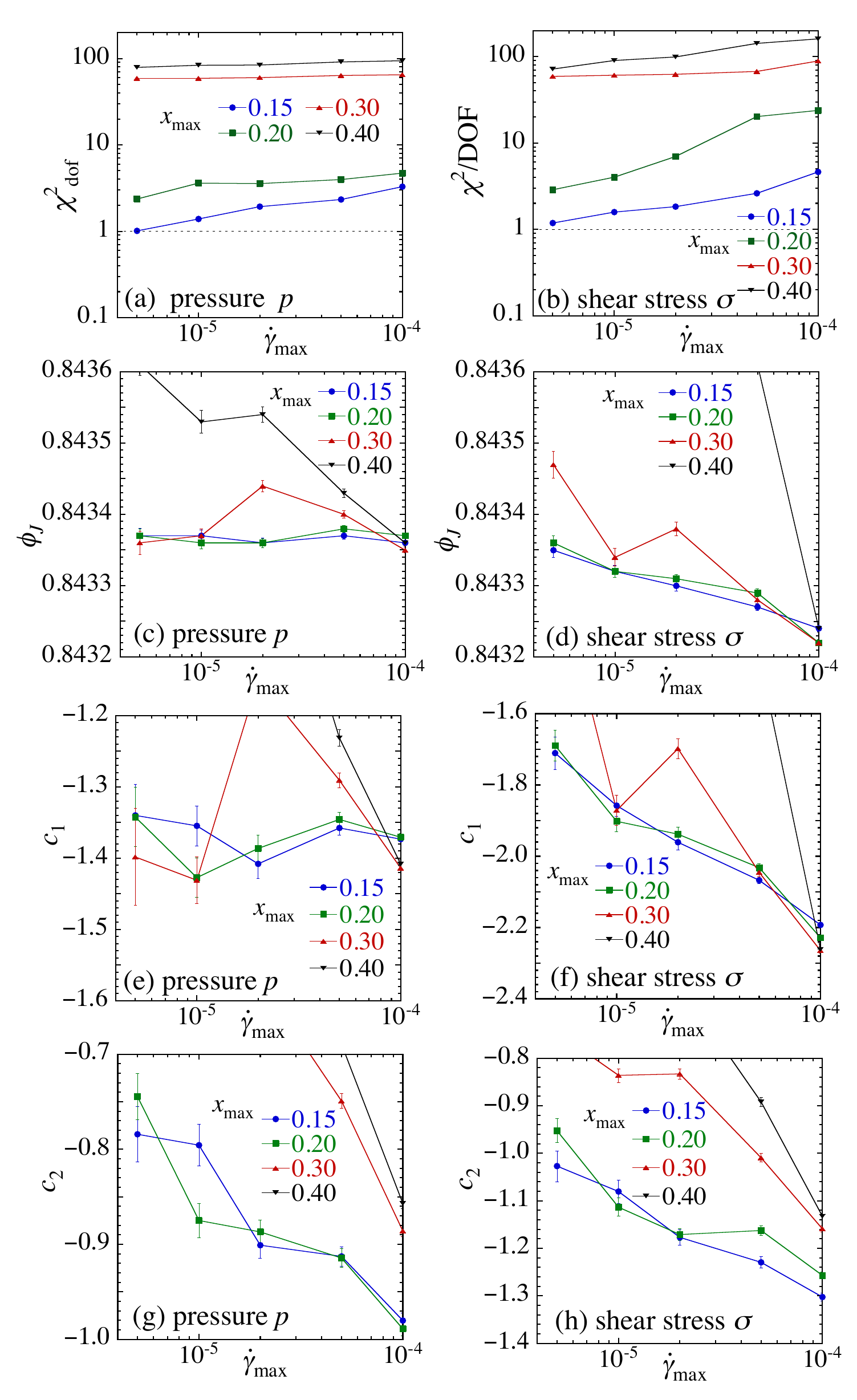}
\caption{(Color online)  Fits of pressure $p$ (left column) and shear stress $\sigma$ (right column) to the scaling form of Eq.~(\ref{eA2}) including presumed logarithmic corrections to scaling.  Here we fix the exponent value $\beta=5.7$ as given by the theoretical prediction of DeGiuli et al. \cite{DeGiuli}, and take  $y=1$ as assumed by Otsuki and Hayakawa \cite{OH1,OH2} and as found for static jamming \cite{OHern}. Different curves represent different cutoffs $x_\mathrm{max}$ on the scaling variable; results are plotted vs the cutoff on the strain rate $\dot\gamma_\mathrm{max}$.  We show results for the chi squared per degree of freedom $\chi^2_\mathrm{dof}$, the jamming fraction $\phi_J$, and the exponents $c_1$ and $c_2$ of the logarithmic corrections.  Because the $\chi^2_\mathrm{dof}$ is so poor for $x_\mathrm{max}>0.2$ (note the logarithmic scale on the vertical axis in panels a and b), in subsequent panels we focus only on the data for $x_\mathrm{max}\le 0.2$.
}
\label{fA4}
\end{figure}

Next we consider the theoretical predictions of DeGiuli et al. \cite{DeGiuli} which give $\beta\approx 5.7$.  Since DeGiuli et al. deal with a hard-core model, they can make no direct prediction about the other exponents.  However if we assume $y=1$ for the harmonic soft-core interaction, as assumed by Otsuki and Hayakawa \cite{OH1,OH2} and as believed to be the case for mechanically stable configurations compressed above the static jamming transition \cite{OHern} (though not consistent with the result we claim in this work), we then have $q=2y/(\beta+y)\approx 0.3$ and $1/z\nu=2/(\beta+y)\approx0.3$.  
Fixing $q$ and $1/z\nu$ to these values we proceed as before, letting all other parameters vary in our fit.  In Fig.~\ref{fA4} we present the resulting values of $\chi^2_\mathrm{dof}$, $\phi_J$, $c_1$ and $c_2$, as functions of the cutoffs $\dot\gamma_\mathrm{max}$ and $x_\mathrm{max}$.  We show results from fits both to pressure (left column) and to shear stress (right column).

We see that the $\chi^2_\mathrm{dof}$ is generally too big to consider these fits to be reasonable (note the logarithmic scale on the vertical axes of Figs.~\ref{fA4}a and b).  Only for the smallest $x_\mathrm{max}=0.15$, $0.20$, at the smaller $\dot\gamma_\mathrm{max}$ might one consider the $\chi^2_\mathrm{dof}$ as reasonable.  In the subsequent panels, therefore, we focus on the results for these two smallest values of $x_\mathrm{max}$ (data for the larger $x_\mathrm{max}$ are thus often falling outside the range of plot).  We see that, as desired, $\phi_J$ and $c_1$ found from the pressure (panels c and e) are roughly independent of $\dot\gamma_\mathrm{max}$ for $x_\mathrm{max}=0.15, 0.20$.  However, this is not the case for the other quantities that, as $\dot\gamma_\mathrm{max}$ decreases, vary over a range considerably larger than the estimated errors on the data points.  Moreover, comparing the values of $c_1$ and $c_2$ found from the pressure with those found from the shear stress, we see that these values span almost non-overlapping ranges instead of being equal.  We conclude that adding the logarithmic corrections of Eq.~(\ref{eA2}) into our scaling analysis does {\em not} by itself make our results consistent with the predictions of DeGiuli et al. \cite{DeGiuli}.

Comparing the $\chi^2_\mathrm{dof}$ of Figs.~\ref{fA3}a,b with that of Figs.~\ref{fA4}a,b, one might be tempted to conclude that the Otsuki-Hayakawa prediction \cite{OH1,OH2} better fits the data than does that of DeGiuli et al. \cite{DeGiuli}.  However it is important to note that none of the fits in this appendix are doing particularly better than the fits of Sec.~\ref{ssWithout};  both the earlier fits of Sec.~\ref{ssWithout} and the fits of this appendix find critical parameters that noticeably vary as one varies the window of data used in the fit, and thus are not providing self-consistent results.
Our results therefore seem better explained by the corrections-to-scaling that arise from the leading irrelevant variable, as discussed in Sec.~\ref{ssCorrections}.

We have also tried fits to Eq.~(\ref{eA2}) assuming slightly different fixed values of $\beta$ and $y$, as well as fits in which only $\beta$ is fixed and $y$ may vary (and vice versa), however we do not find results that are any more satisfactory. A more accurate test for the presence of logarithmic scaling corrections would depend on knowing precise values for the leading exponents $q$ and $1/z\nu$ (or equivalently $\beta$ and $y$), but unfortunately these are not known.   
While our results therefore cannot rule out the presence of logarithmic corrections, neither do they give any support for them.  Our results do not rule out the possibility that $d_{uc}=2$, however they do show that the addition of logarithmic corrections alone is not sufficient to make our data compatible with either of the two theoretical predictions in \cite{OH1,OH2} or \cite{DeGiuli} for the leading critical exponents.

As a final comment we note that if indeed $d_{uc}=2$, one would expect to see scaling with mean-field exponents with {\em no} logarithmic corrections if one carried out simulations in $d=3>d_{uc}$ dimensions.  To obtain sufficiently accurate data for our model in $d=3$ is a computationally challenging project that we leave for future investigation.  However we may note that $d=3$ simulations have been carried out by Kawasaki et al. \cite{Kawasaki} for a simpler model with Newtonian rheology.  In that case they found (as we similarly found \cite{OlssonTeitelScaling} for this Newtonian model in $d=2$) that a simple scaling analysis as in Sec.~\ref{ssWithout} cannot explain the shear stress over the range of strain rates $\dot\gamma$ studied;  their subsequent analysis is equivalent to the correction-to-scaling approach described here in Sec.~\ref{ssCorrections}, and as used by us \cite{OlssonTeitelScaling} to explain results for this Newtonian model in $d=2$.  We therefore might expect that similar corrections-to-scaling terms, from the leading irrelevant variable, would be present in the present model even in $d=3$, and so presumably also in $d=2$, as we argue is the case in this work.


\begin{thebibliography}{99}
\bibitem{xmax}Since one needs to know the values of $\phi_J$ and $1/z\nu$ to determine the scaling variable $x$, we make an initial guess for these values, select data according to these initial values, then do a fit to obtain new values for $\phi_J$ and $1/z\nu$.  We iterate this procedure until we get consistent results.
\end{thebibliography}
\end{document}